\documentclass[preprint2,tighten]{aastex}
\usepackage{epsfig}

\def\cha{{\it Chandra\/}} 

\def\mathnew{\mathsurround=0pt}
\def\simov#1#2{\lower .5pt\vbox{\baselineskip0pt \lineskip-.5pt
\ialign{$\mathnew#1\hfil##\hfil$\crcr#2\crcr\sim\crcr}}}
\def\simgreat{\mathrel{\mathpalette\simov >}}
\def\simless{\mathrel{\mathpalette\simov <}}
\def\arcmin{\hbox{$^\prime$}}
\def\arcsec{\hbox{$^{\prime\prime}$}}

\def\xmm{{\it XMM-Newton\/}}
\def\MeV{Me\kern-0.11em V}
\def\keV{ke\kern-0.11em V}

\slugcomment{ }

\shorttitle{Fossil Group at z = 0.59}
\shortauthors{Ulmer et al.}

\begin{document}


\title{Cl 1205+44, a fossil group at \em{z} = 0.59}


\author{\vspace*{-10pt}M. P. Ulmer} 
\affil{Department Physics \& Astronomy, Northwestern University,
    Evanston, IL 60208-2900}
\email{m-ulmer2@northwestern.edu}

\author{C. Adami,~~ G. Covone}
\affil{Laboratoire d'Astrophysique de Marseille, Traverse du Siphon,
13012 Marseille, France}
\email{christophe.adami@oamp.fr,~~ giovanni.covone@oamp.fr}

\author{F. Durret}
\affil{Institut d'Astrophysiqe de Paris, CNRS, 98 bis Boulevard Arago, 75014
Pars, France}
\email{durret@iap.fr}

\author{G. B. Lima Neto }
\affil{Instituto de Astronomia, Geof\'{\i}sica e C. Atmosf./USP, R. do Mat\~ao
1226,
05508-090 S\~ao Paulo/SP, Brazil}
\email{gastao@astro.iag.usp.br}

\author{K. Sabirli\altaffilmark{1}}
\affil{Carnegie Mellon University, 5000 Forbes Avenue, Pittsburgh, PA 15213, USA}
\email{sabirli@sussex.ac.uk}

\author{B. Holden\altaffilmark{2}}
\affil{Department of Physics, University of California, Davis, 1
Shields Avenue, Davis, CA 95616}
\email{holden@ucolick.org}

\author{R.G. Kron}
\affil{Fermi National Accelerator Laboratory, MS 127, Box 500, Batavia, IL 60510}
\email{rich@oddjob.uchicago.edu}

\author{A.K. Romer\altaffilmark{1}}
\affil{Carnegie Mellon University, 5000 Forbes Avenue, Pittsburgh, PA 15213, USA}
\email{romer@sussex.ac.uk\vspace*{-30pt}}

\altaffiltext{1}{present address: Astronomy Center, Department of Physics and Astronomy, 
University of Sussex, Falmer, Brighton, BN1 9QH}
\altaffiltext{2}{present address: UCO/Lick Observatories, 1156 High Street, Santa
Cruz, CA 95065}

\begin{abstract}

This is a report of Chandra, \xmm, HST and ARC observations of an
extended X-ray source at z = 0.59. The apparent member galaxies range
from spiral to elliptical and are all relatively red (i$'$-$K_{s}$ about
3). We interpret this object to be a fossil group based on the
difference between the brightness of the first and second brightest
cluster members in the $i'$-band, and because the rest-frame
bolometric X-ray luminosity is about $9.2 \times 10^{43}$
h$_{70}^{-2}$ erg s$^{-1}$. This makes Cl 1205+44 the highest redshift
fossil group yet reported.  The system also contains a central
double-lobed radio galaxy which appears to be growing via the
accretion of smaller galaxies.  We discuss the formation and evolution
of fossil groups in light of the high redshift of Cl 1205+44.

\end{abstract}


\section{Introduction}  In a ROSAT survey of extended sources
\citep{Adami2000,Romer2001}, we found a source showing extended X-ray
emission but with only a single faint optical (R-band) counterpart. On
deep optical images, the colors of the galaxies in the X-ray area were
red enough that the program HYPERZ derived a photometric redshift
greater than 1.  This led to Chandra and \xmm\ observations of the
object we designate as Cl~1205+44.  We find that the system is a fossil
group at the highest redshift yet published (0.59 versus typical
values of $ \simless 0.1$).  The system is particularly interesting
because it allows us to explore evolutionary tracks of fossil groups
and to consider scenarios for the evolution of galaxies in groups.

Fossil groups have been defined \cite[e.g.][]{Jones2003} as being
similar to groups (poor clusters) in which the luminosity function of
the member galaxies has been modified by accretion of galaxies onto
the central galaxy while the rest of the system remains unevolved for
approximately 4 Gyrs.  The galaxy accretion onto the D galaxy leads to
a magnitude difference between the first and second brightest galaxies
(m$_{12}$) of 2 or higher in the rest frame R or V bands within one
half a virial radius, where \cite{Jones2003} use the following to
calculate the virial radius: r$_{vir}$ = $ 3.89\times (T/10$
\keV)$^{0.5} \times (1+z)^{-3/2}$ h$_{50}^{-1}$ Mpc.

Moreover, Cl 1205+44 is interesting because it harbors a double lobed
radio source which is also a D (central dominant) galaxy. Such systems
can be directly compared with the model of \cite{West1994}.  Also, the
radio source simply makes the group more complex and its formation
could be related to cooling flows\footnote{Cooling flow literature is
too extensive to comprehensively reference; a few references that
directly relate radio lobes to X-rays in relatively poor systems with
X-ray emission are: \cite{Carilli1994,Harris2000,McNamara2000}} on
one hand, and energy injection on the other \cite[e.g.][]{Sun2003}.

A D galaxy that is a radio source is a tracer for finding fossil groups which
led J.~S. Mulchaey (2003, private communication; see also
\citeauthor{Mulchaey2003}, \citeyear{Mulchaey2003}) independently to identify
this object as a fossil group and to obtain HST and redshift observations. The
HST and redshift data nicely complement the \cha, \xmm, $i'$- and $K_{s}$-band data
we have obtained.

Our study of Cl~1205+44 also adds one more data point to the topic of
preheating. Recent studies of simulations of preheating can be found,
for example, in \cite{Borgani2004}.  The possibility of preheating of
the ICM in groups and clusters has taken on added significance, as it
relates to the Suyanev-Zel'dovich (SZ) effect, which can influence the
interpretation of the high order power spectrum of the Cosmic
Microwave Background \cite[CMB;][]{Lin2004}.

In this paper, we report the results of our multi-wavelength analysis and
discuss how Cl 1205+44 fits into the larger picture of fossil group formation
and evolution.

We will use H$_0$=70 km s$^{-1}$ Mpc$^{-1}$, $\Omega_\Lambda$=0.7 and
$\Omega_m$=0.3 hereafter.  At a redshift of 0.5915, which is the most likely
redshift of Cl 1205+44, the angular scale for this cosmology is 6.64
kpc/\arcsec.

\section{Observations and Analysis}

\subsection{Optical and IR observations}

We made observations in the $i'$ and $K_{s}$-bands at the ARC
telescope\footnote{See http://www.apo.nmsu.edu for details.}. For the
$i'$-band, we used the Spicam camera for a net total exposure time of
7,200 seconds at a mean airmass of 1.15, and for the $K_{s}$-band we used
the GRIMII camera for a net total exposure time of 7,200 seconds. We
took separate 10 minute exposures to acquire the $i'$-band data and 10
to 30 second exposures to accumulate the $K_{s}$ data.  The $i'$-band data
were reduced using the ESO-MIDAS package and the $K_{s}$ data were reduced
using DIMSUM, a NOAO/IRAF tool.  The zero points were computed using
standard stars observed at the same time as the scientific data and at
similar air masses.

The HST data were retrieved from the archive (proposal ID: 8131, PI:
Mulchaey).  They consist of $6 \times 1200$ second dithered R-band
WFPC2 (F702W filter) exposures. We reduced the data using the {\em
drizzle} IRAF package \citep{Fruchter2002}. We generated catalogs of
detected objects using SExtractor \cite[version 2.3][]{Bertin1996}
using a detection threshold of 1.5$\sigma$ and an analysis threshold
of 2$\sigma$.

\begin{figure}
\begin{center}
\centerline{\epsfig{figure=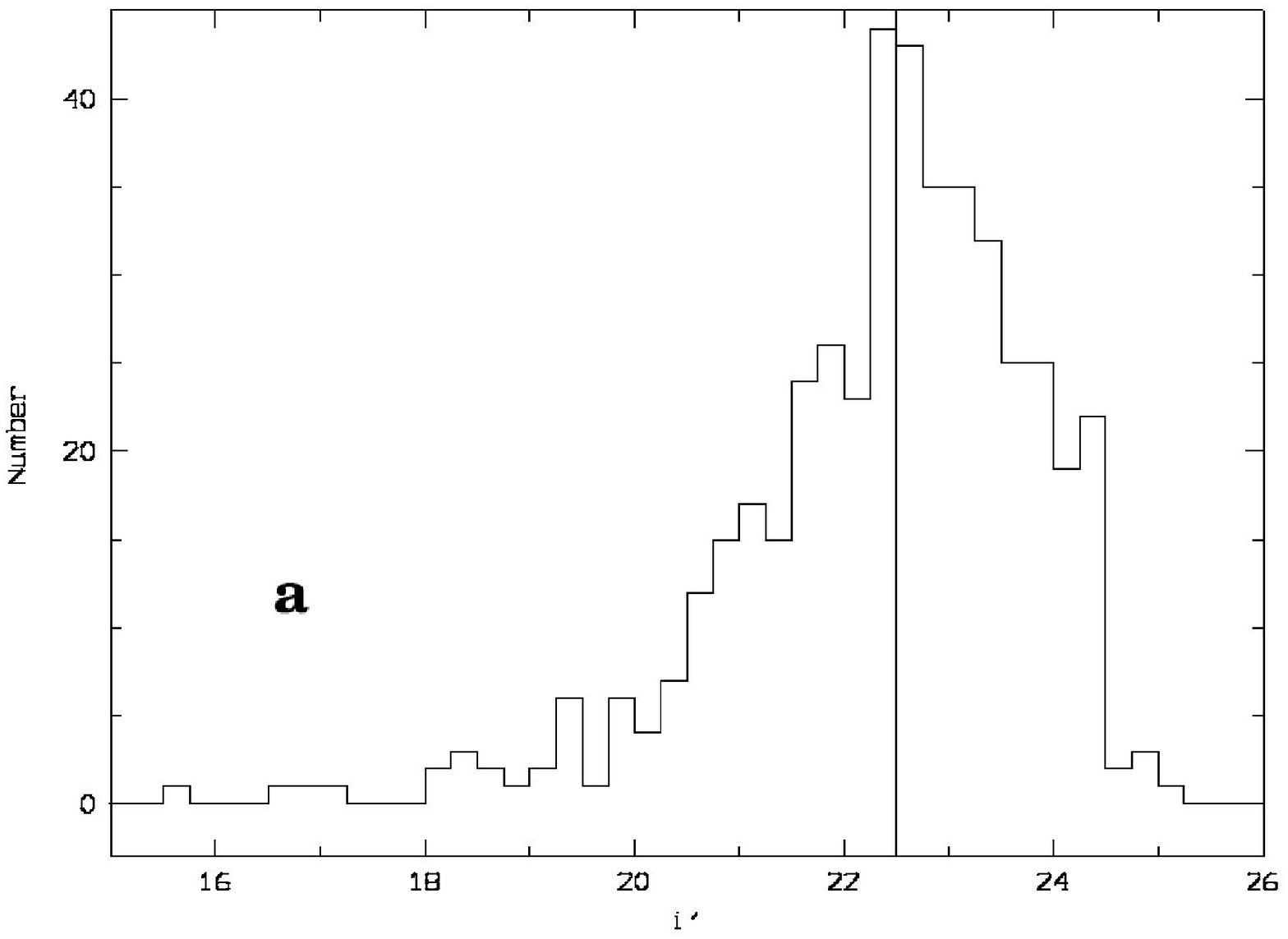,width=2.8in,angle=0}}
\vspace{0.1in}
\centerline{\epsfig{figure=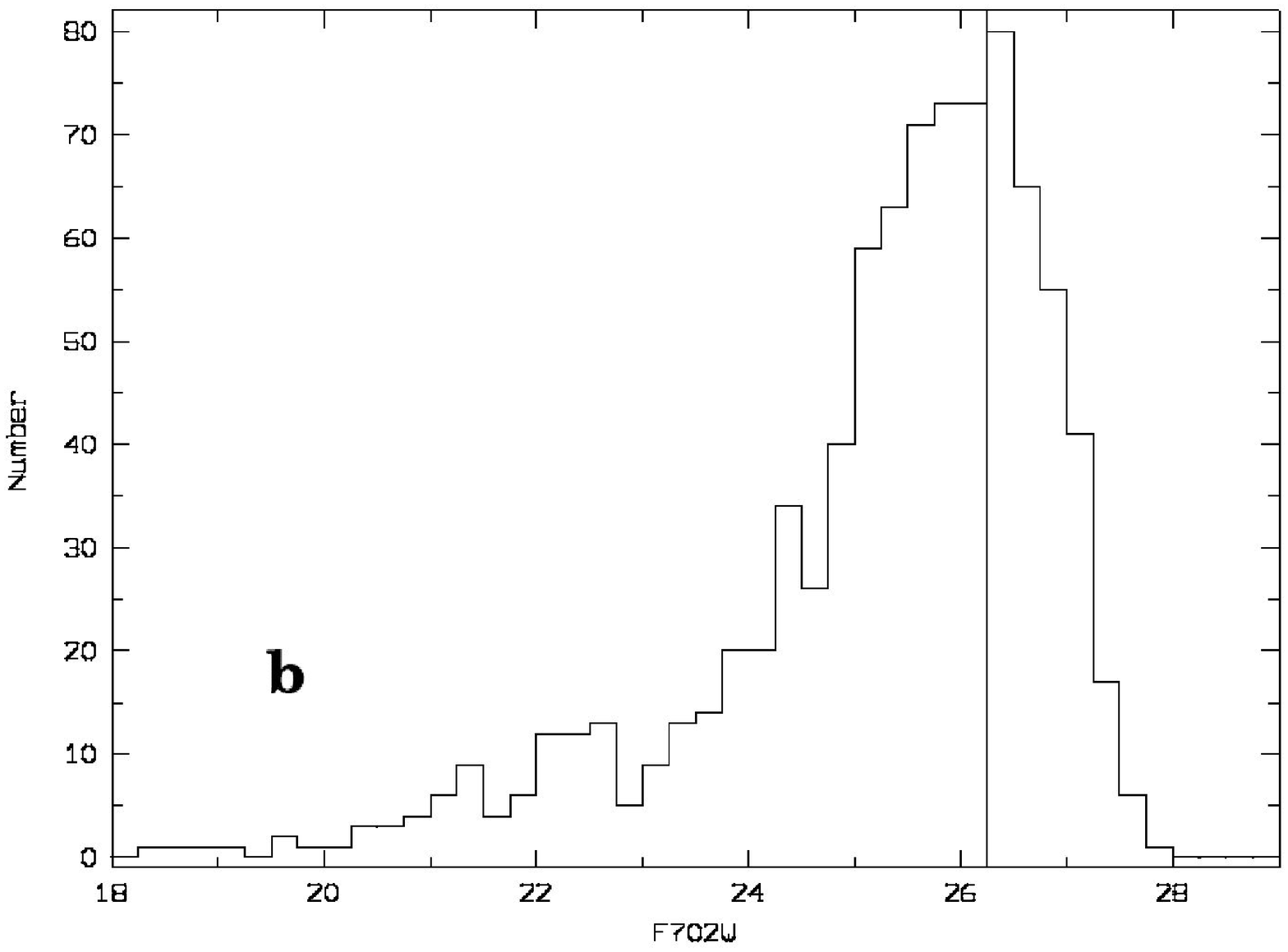,width=2.8in,angle=0}}
\vspace{0.1in}
\centerline{\epsfig{figure=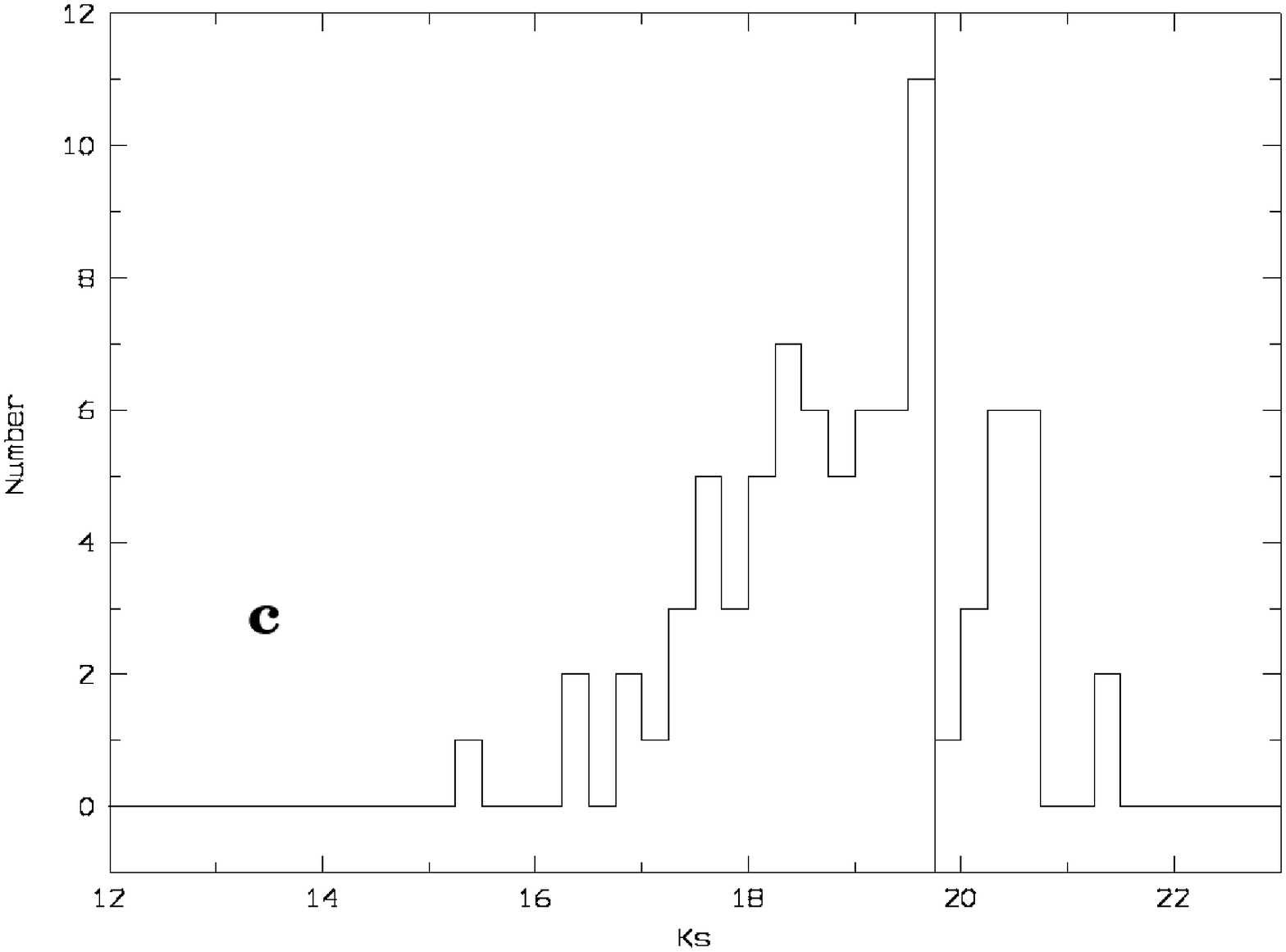,width=2.8in,angle=0}}
\caption{\label{hist}Histograms of the total magnitudes in the ARC
Spicam field of view for $i'$, in the ARC GRIMII field of view for $K_{s}$
and in the HST WFPC2 field of view for F702W. The vertical lines 
give the upper limits for the completeness: $i'$$\sim$22.25,
F702W$\sim$26.25 and $K_{s}$$\sim$19.75.}
\end{center}
\end{figure}

We used the fixed aperture $3''$ magnitudes from SExtractor for the HST
magnitudes and we used the SExtractor auto-magnitude feature to derive the 
magnitudes for the ARC data.  Figure~\ref{hist} gives the magnitude histograms
in the three magnitude bands available to us (in the ARC Spicam field of view
for $i'$, in the ARC GRIMII field of view for $K_{s}$ and in the HST WFPC2 field of
view for F702W). These histograms allow us to estimate an upper limit to the
completeness value of the magnitude in each bands: $i'$$\sim$22.25,
F702W$\sim$26.25 and $K_{s}$$\sim$19.75.  The respective solid angle coverage of the
instruments is: WFPC2 image: 2.2$\times$2.2~arcmin$^2$, $i'$ image:
6$\times$4.3~arcmin$^2$, $K_{s}$ image: 2$\times$2~arcmin$^2$.

\begin{figure}
\begin{center}
\centerline{\epsfig{figure=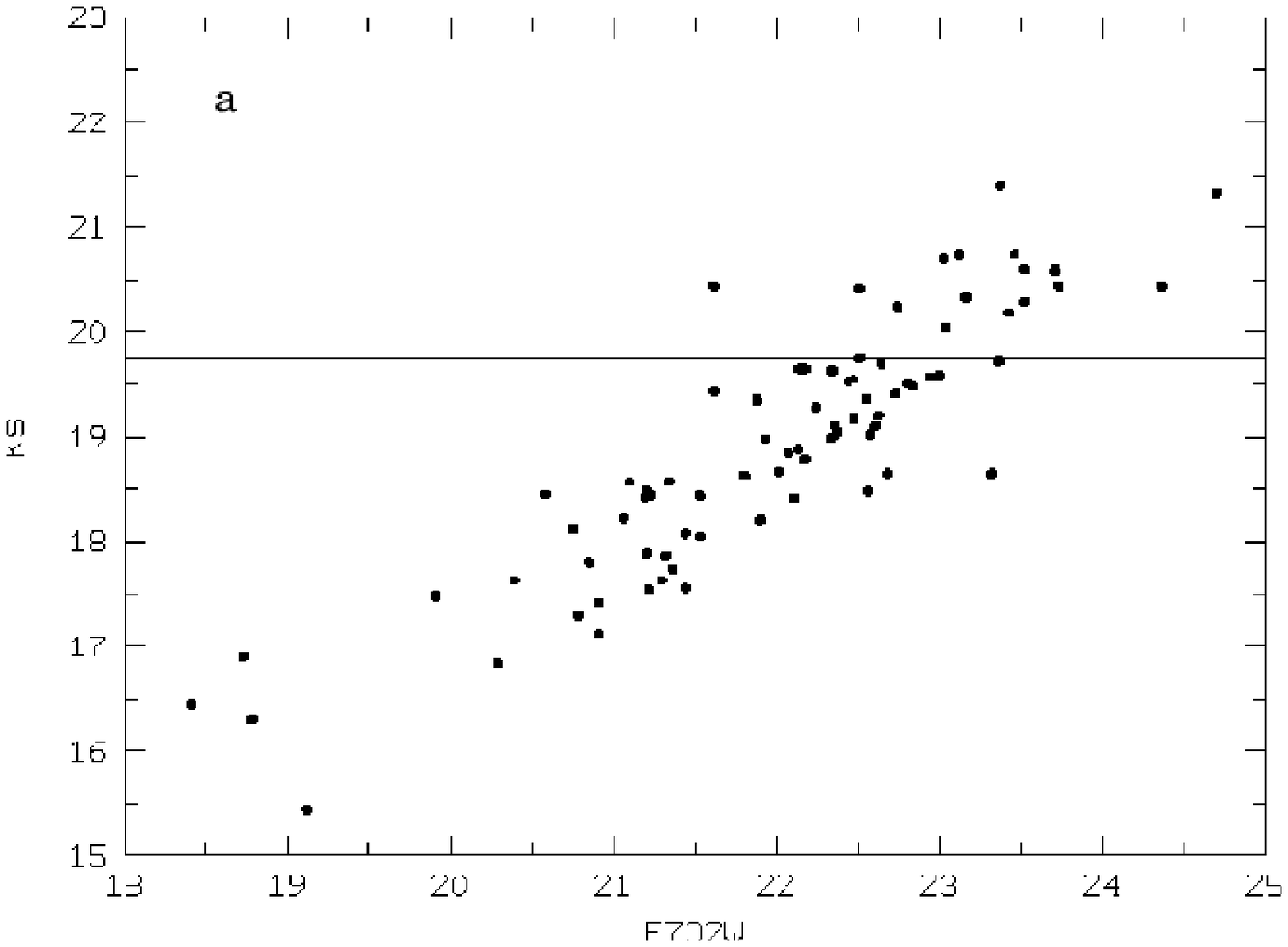,width=2.8in,angle=0}}
\vspace{0.1in}
\centerline{\epsfig{figure=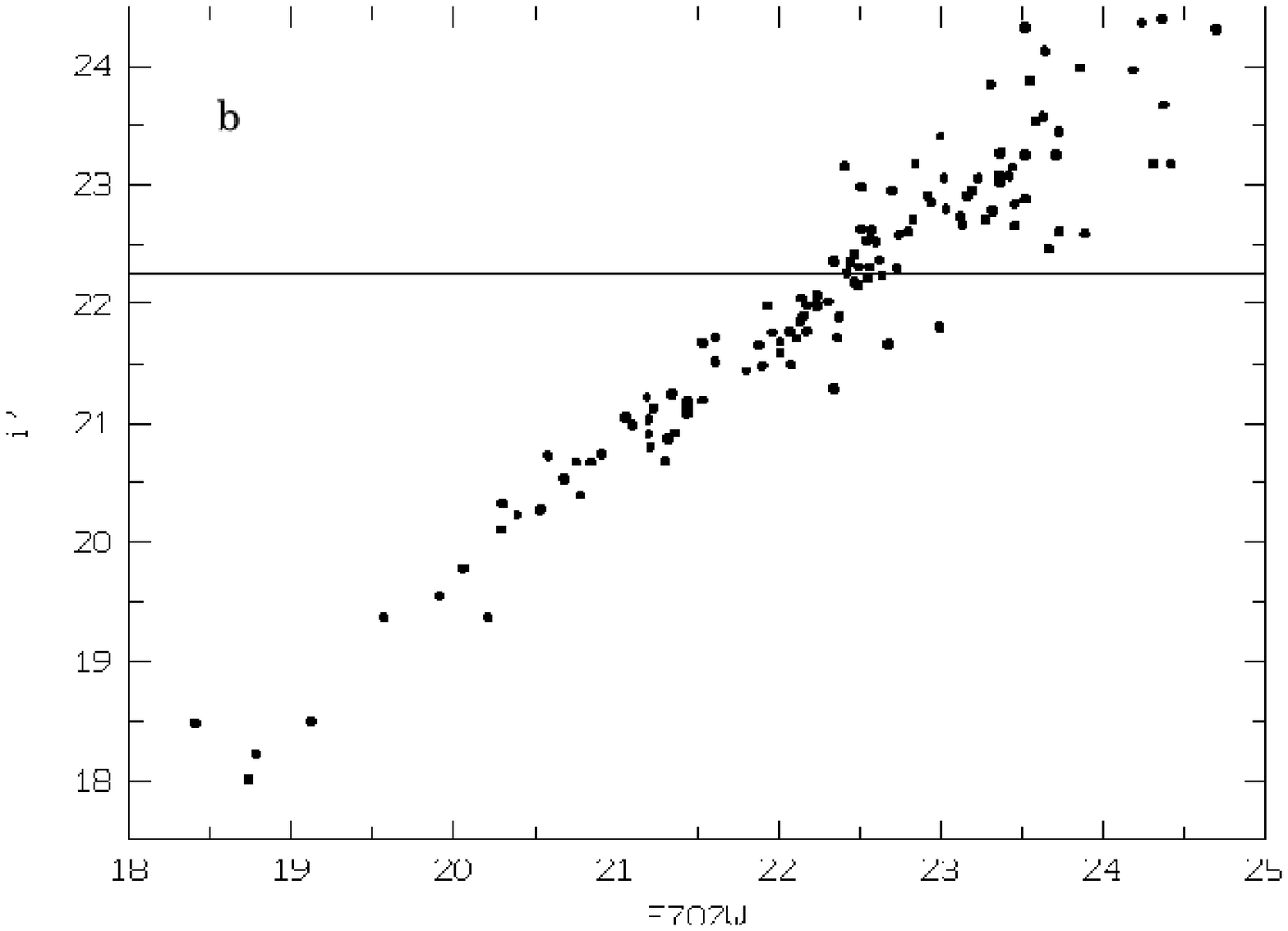,width=2.8in,angle=0}}
\vspace{0.1in}
\centerline{\epsfig{figure=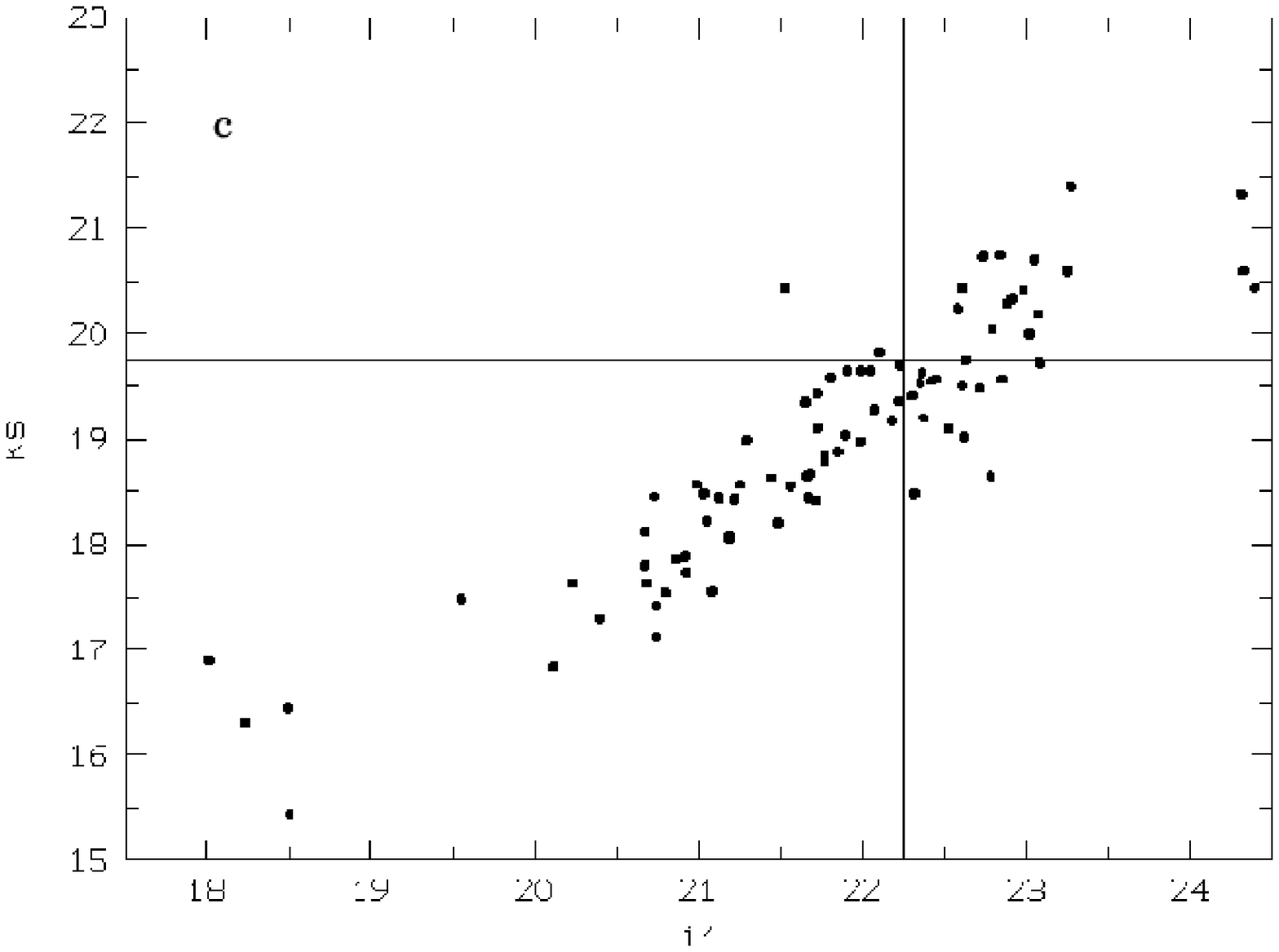,width=2.8in,angle=0}}
\caption{\label{colours}Relation between our three magnitudes plotted against
each other two at a time.
The horizontal lines and vertical line show the completeness levels for the
i' and $K_{s}$ data taken from in Figure \ref{hist}.
}
\end{center}
\end{figure}

In order to assess the quality of our data, we plotted the magnitudes in the
different bands against each other as shown in  Figure~\ref{colours}. We
estimate from the results a relative 1$\sigma$ uncertainty between magnitudes
(limiting ourselves to the brightest completeness limit) of 0.26 mag between
F702W and $i'$ and of 0.36 mag between $i'$ and $K_{s}$. This gives an upper limit
to the magnitude uncertainty since some of the scatter is due to the
intrinsic color variations of the objects. 

As an external test, we compared our $K_{s}$ magnitudes with the estimates
for the two objects in the $K_{s}$ field bright enough to be detected by
2MASS\footnote{http://www.ipac.caltech.edu/2mass/} (including the D
galaxy). We have reasonable agreement (for the two objects,
$K_{s}$$-$K$_{2MASS}$=$-0.38$ and +0.26) given that these objects are below
the 2Mass completeness limit.

\begin{figure*}  
\begin{center} 
\vspace{-0.6 in}
\parbox{4.2in}{\epsfig{figure=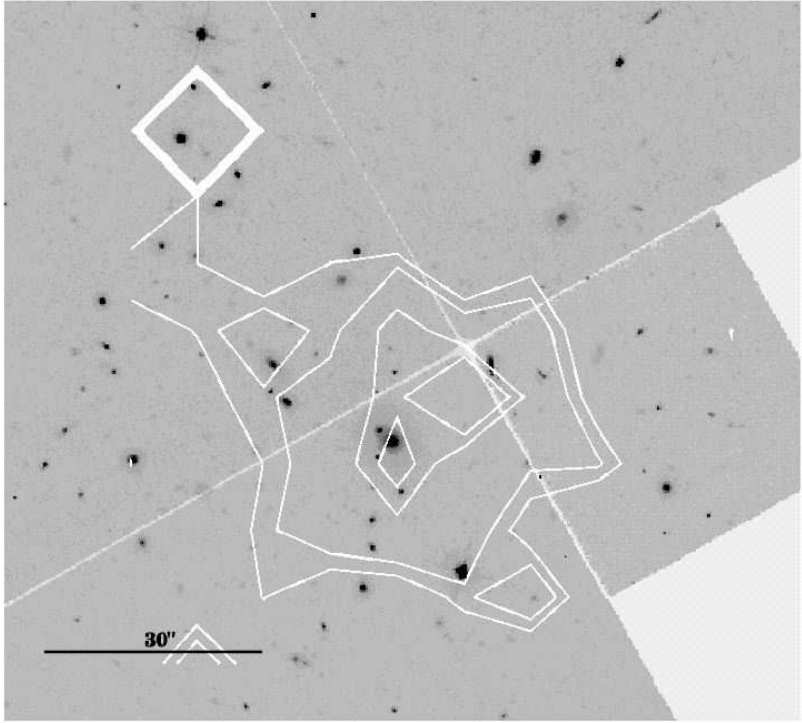,width=4.2in}} 
\caption{\label{hstchandra}Chandra
contours from the 0.5--8 keV image superimposed on the F702W HST image of
Cl~1205+44.  We used a
smoothing window for the contours of 4.\arcsec5, which is a finer  scale than
we used in Figures \ref{firsthst} and \ref{firstcha}. North is up and East 
to the left.}  
\end{center} 
\end{figure*}

\begin{figure}
\plotone{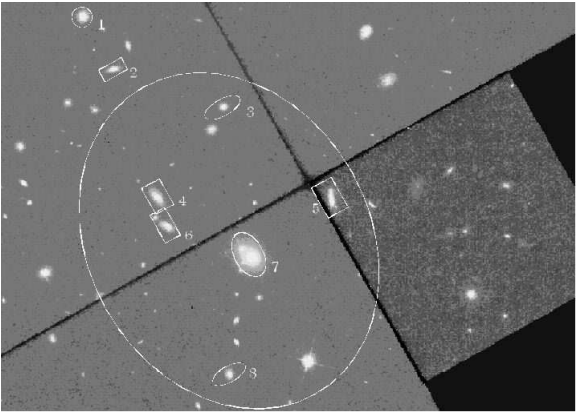}
\caption{\label{melmarked}An expanded image of the HST $\sim$ 0.\arcmin97 
$\times$ 0.\arcmin72 field. North is up and east to the left.  The northernmost
circled galaxy is likely to be a line-of-sight AGN. The galaxies marked with
rectangles are late types and those with ellipses are early types. The large
ellipse represents the approximate extent of the X-ray emission, see Figure
\ref{hstchandra} for a more accurate superposition of the X-ray flux onto the
HST image. Starting from the north the respective 
Id numbers to correlate with Table~\ref{galaxies} i$'$ magnitudes and 
($K_{s}$$-i'$) values are: \#1,18.23(1.93), \#2 20.80 (3.26),\#3 21.19 (3.14), 
\#4 20.43 (3.14),\#5
20.65 (3.02), \#6 20.99 (2.42), \#7 18.50 (3.06), \#8 20.85 (2.99).}

\end{figure}

For use in our discussion below, we calculated the total R-band luminosity. To
derive this number, we used the total F702W flux enclosed\footnote{We excluded 
the star, which is the brightest object toward the top of the figure but inside
the contours; it is south-southwest of the D galaxy} by the X-ray contours in 
Figure \ref{hstchandra}. We estimate that approximately 90\% of our derived
value of L$_R$ comes from the galaxies enclosed by rectangles in Figure
\ref{melmarked}. Therefore, the L$_R$ value is probably accurate to within
20\%. The 20\% uncertainly is also consistent with subtracting the flux of \#6 
in Figure \ref{hstchandra} (see also Table \ref{galaxies}) which, based on its
colors, is the most likely in our list not to be a cluster member. The
uncertainty in L$_R$ does not affect any conclusion we draw in the discussion
section of this paper. The F702W total luminosity is $6.0 \times 10^{11}$
L$_{\bigodot}$. Converting to the R$_J$ band used by \cite{Jones2003}, we
obtain L$_R$ = $5.5\times 10^{11}$ L$_{\bigodot}$.  The K-correction to {\em z}
= 0 gives a (rest frame) value of L$_R = 7.8 \times 10^{11}$ L$_{\bigodot}$.
Finally, to compare with \citeauthor{Jones2003}, we  use H$_0$=50 and q$_0$ =
0.5 which leads to L$_R = 8.9 \times 10^{11}$ L$_{\bigodot}$.

\subsection{Redshift} \label{redshift} Since the X-ray data were not of
sufficient signal to noise ratio to constrain the cluster redshift (see below),
we needed optical spectra of the galaxies belonging to the system. By chance,
this system was observed by J.~S. Mulchaey (2003, private communication)
in his ongoing fossil group survey and he kindly provided us with the value of
the redshift of the brightest galaxy of the putative fossil group, {\em z} =
0.5915. These data plus the positional coincidence of the brightest group
galaxy  (called the D galaxy hereafter) with the centroid (see \S \ref{xrlum})
of the X-ray emission firmly identifies the extended X-ray emission with the
brightest galaxy and the associated galaxies in Table~\ref{galaxies}.

\subsection{X-ray Observations}

We were granted time to observe Cl 1205+44 with the \xmm\ (June 2003,
52,200~s) and \cha\ (October 2003, 31800~s) satellites.

\subsubsection{Chandra data}

The Chandra observation was made in ``Very Faint'' mode with a time
resolution of 3.24~sec and a CCD temperature of $-120^{\circ}$C. The
data were reduced using CIAO version
3.0.1\footnote{\texttt{http://asc.harvard.edu/ciao/}} following the
Standard Data Processing, producing new level 1 and 2 event files.

We have further filtered the level 2 event file, keeping only \textit{ASCA}
grades 0, 2, 3, 4 and 6, and restricted our data reduction and analysis to the
back-illuminated chip, ACIS-S3. We checked that no afterglow was present and
applied the Good Time Intervals (GTI) supplied by the pipeline. Then, we have
checked for flares using the light-curve in the [10--12 keV] band; no flare
was detected and the total exposure time was 29,711~s.

We have used the CTI-corrected ACIS background event files (``blank-sky''),
produced by the ACIS calibration
team\footnote{\texttt{http://cxc.harvard.edu/cal/Acis/WWWacis\_cal.html}},
available from the calibration data base (CALDB). The background events were 
filtered, keeping the same grades as the source events, and then were 
reprojected to match the sky coordinates of the Cl 1205+44 ACIS observation.

\subsubsection{\xmm\ data}

Cl~1205+44 was observed in standard Full Frame mode using the ``thin''
filter with the two EPIC MOS1 and MOS2 and the PN detectors. The basic
data processing (the ``pipeline'' removal of bad pixels, electronic
noise and correction for charge transfer losses) was done with package
SAS V5.3, thus creating calibrated event files for each detector.

For the MOS1 and MOS2 cameras, following the standard procedure, we have
discarded the events with FLAG~$\neq 0$ and PATTERN~$> 12$; for the PN, we have
restricted the analysis to the events with PATTERN $\le$ 4 and Flag =0.

The light-curves in the [10--12 keV] band that we have produced showed
that there were severe flaring events during the observation. Filtering out
the periods with flares substantially reduced the exposure times:
21,223~s, 20,861~s, and 16,478~s for the MOS1, MOS2 and PN,
respectively (from an initial 52.2~ks).

With the cleaned event files, we have created the redistribution
matrix file (RMF) and ancillary response file (ARF) with the SAS tasks
\texttt{rmfgen} and \texttt{arfgen} for each camera and for each
region that we have analyzed.

The background was taken into account by extracting spectra  from the blank sky
templates described by \cite{Lumb2002}, and reprojected to the same
coordinates and roll angle of the Cl~1205+44 \xmm\ observation. The same
filtering procedure was applied to the background event files.  We give a
breakdown of the number of counts used per detector in the fitting in
Table~\ref{counts}.

\subsubsection{X-ray Spectral fits}

Spectra were analyzed with XSPEC~11.3. We have simultaneously fit all
four spectra from the \xmm\ MOS1, MOS2, PN and Chandra ACIS-S3
cameras. The spectra were rebinned so that we could use the standard
$\chi^{2}$ minimization when fitting the spectra. We have applied the
\textsc{mekal} plasma spectral model  \citep{Kaastra93,Liedahl95}.

\begin{figure}
\plotone{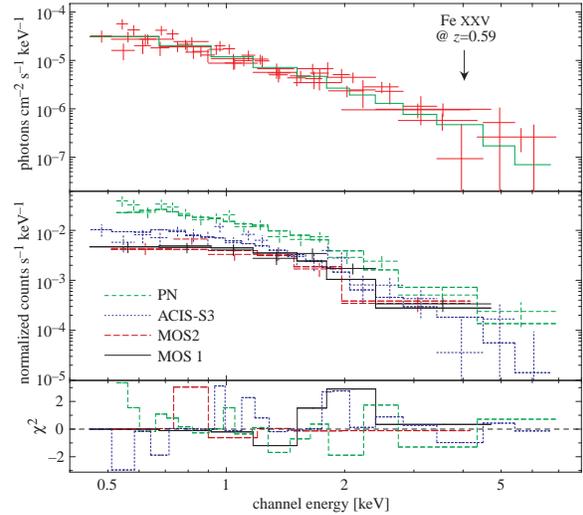}
\caption{\label{MEKAL}\textsf{Top:} Flux in photons~cm$^{-2}$~s$^{-1}$~keV$^{-1}$, with
data points from all cameras. For simplicity in the black and white version, 
there is only one line drawn of the best fit spectrum. Except for the lowest
energy bin which is dominated by the MOS data, the fit is dominated by the PN
data. The redshifted Fe\textsc{XXV} line for $z = 0.59$ is indicated.
\textsf{Middle:} Fluxes in counts~s$^{-1}$~keV$^{-1}$ for all cameras: full
line, MOS1; long dashed line, MOS2; dotted line, PN; short dashed line,
ACIS-S3. The differences among best fit spectra come from the different
detector responses. \textsf{Bottom:} Residues given as the $\chi^{2}$
contribution of each energy bin. See the electronic version of the Journal for
a color version of this figure.}
\end{figure}

\begin{figure}
\plotone{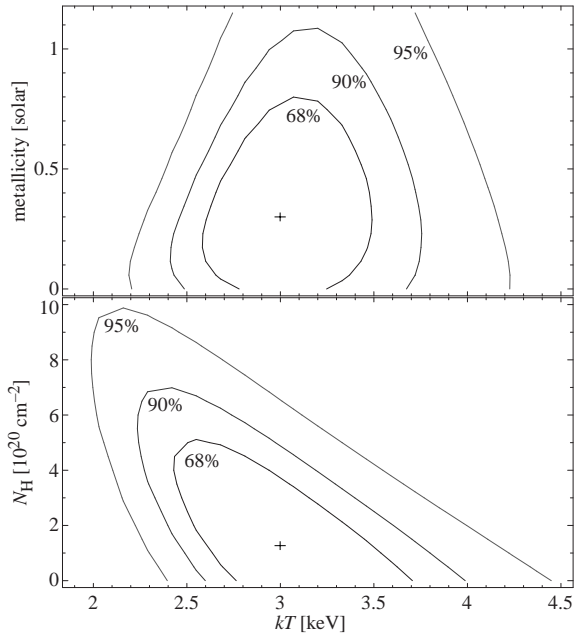}
\caption{\label{contours} Probability contours of metallicity and N$_H$ versus
kT fits to the X-ray data, see text.}
\end{figure}

The photoelectric absorption (mainly due to neutral hydrogen in our
galaxy) was computed using the cross-sections given by
\cite{Balucinska92}, available in \textsc{xspec}. Given the low
count-rate, the hydrogen column density $N_{\rm H}$ could not be
well constrained by the spectral fit, therefore we fixed it to the galactic
value at the cluster position. The interpolation of the HI map of
\citet{Dickey90}, using the task \texttt{nh} from \textsc{ftools},
yields $N_{\rm H} = 1.27 \times 10^{20}\,$cm$^{-2}$.
The results are shown in Figure \ref{MEKAL}.
The metallicity is poorly constrained; as shown in Table~\ref{tbl:mekalfit}, at
a $1 \sigma$ confidence level we found $Z < 0.6 Z_{\bigodot}$ and, at 90\%
confidence level,  $Z < 1.1 Z_{\bigodot}$.  For completeness in
Figure~\ref{contours} we show how  $N_{\rm H}$ and metallicity correlate with
kT in the spectral fits.

We restricted the spectral analysis within the energy range where the
cluster emission was above the background and the detectors were well
calibrated.  For the MOS1 and 2 cameras, this interval was [0.3--8.0
keV], for the PN [0.5--8.0 keV] and for ACIS-S3 [0.4--7.0 keV]. The
spectra were extracted inside a circle of 1.4~arcmin radius, centered
on the cluster. The strong northeast X-ray point source (probably an AGN) is
outside this radius.

\subsubsection{X-ray luminosity} 
\label{xrlum}
For the reasons stated above (\S\ref{redshift}), we assume that the cluster
redshift is {\em z} = 0.5915 and fixed this value from hereon.
Table~\ref{tbl:mekalfit} summarizes the best spectral fits, either fixing the
metallicity or the hydrogen column density or both. Clearly, the metallicity
is not well constrained and we only have upper limits for $N_{\rm H}$. Fixing
the metallicity to the typical value found in clusters
\cite[e.g.][$0.3Z_{\bigodot}$]{Fukazawa2000} and $N_{\rm H}$ to the galactic
value, the mean temperature is $kT = 3.0_{-0.3}^{+0.3}$ keV. This system
therefore has an X-ray temperature typical of a poor cluster (such as Abell
194, which has a temperature of $2.6\pm 0.15$~keV [\citeauthor{Nikogossyan99}
\citeyear{Nikogossyan99}], and the poor cluster RX J0848+4456, which has a
temperature of $3.2\pm 0.3$ keV [\citeauthor{Holden2001},
\citeyear{Holden2001}]) and warmer than that of a typical fossil group
\cite[e.g.][]{Jones2003}.

With redshift {\em z} = 0.5915, the corresponding unabsorbed luminosity and
flux in the [2.0--10.0 keV] are $(3.3 \pm 0.3) h_{70}^{-2} \times
10^{43}\,$erg~s$^{-1}$ and $(1.5 \pm 0.2) \times
10^{-14}\,$erg~s$^{-1}$~cm$^{-2}$. The bolometric luminosity is $(9.2\pm 0.4)
h_{70}^{-2} \times 10^{43}\,$erg~s$^{-1}$. The cluster properties are
summarized in Table \ref{tab:sum}. We define the position of the cluster 
to be that of the D galaxy, as there is a peak in the contours (see
Figure \ref{hstchandra}) at this location.  

\begin{figure}
\plotone{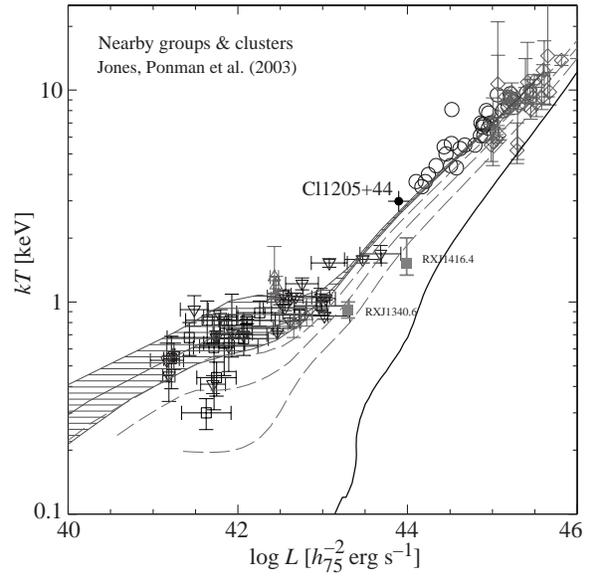}
\caption{\label{jones23}Figure 3 of Jones et al. (2003) with
Cl~1205+44 indicated. The solid lines and cross hatched area are predictions from
\cite{Babul2002} based on an preheating entropy of kTn$_e^{-2/3} \simeq 427$ keV
cm$2$ , and the dashed curves are the predictions for 300, 200, and 100 keV
cm$^{2}$.  The lowest curve best fits the two ROSAT data points for fossil
groups annotated with RXJ names. See Jones et al. for details.}
\end{figure}

\begin{figure*}
\begin{center}
\parbox{4.5 in}{\epsfig{figure=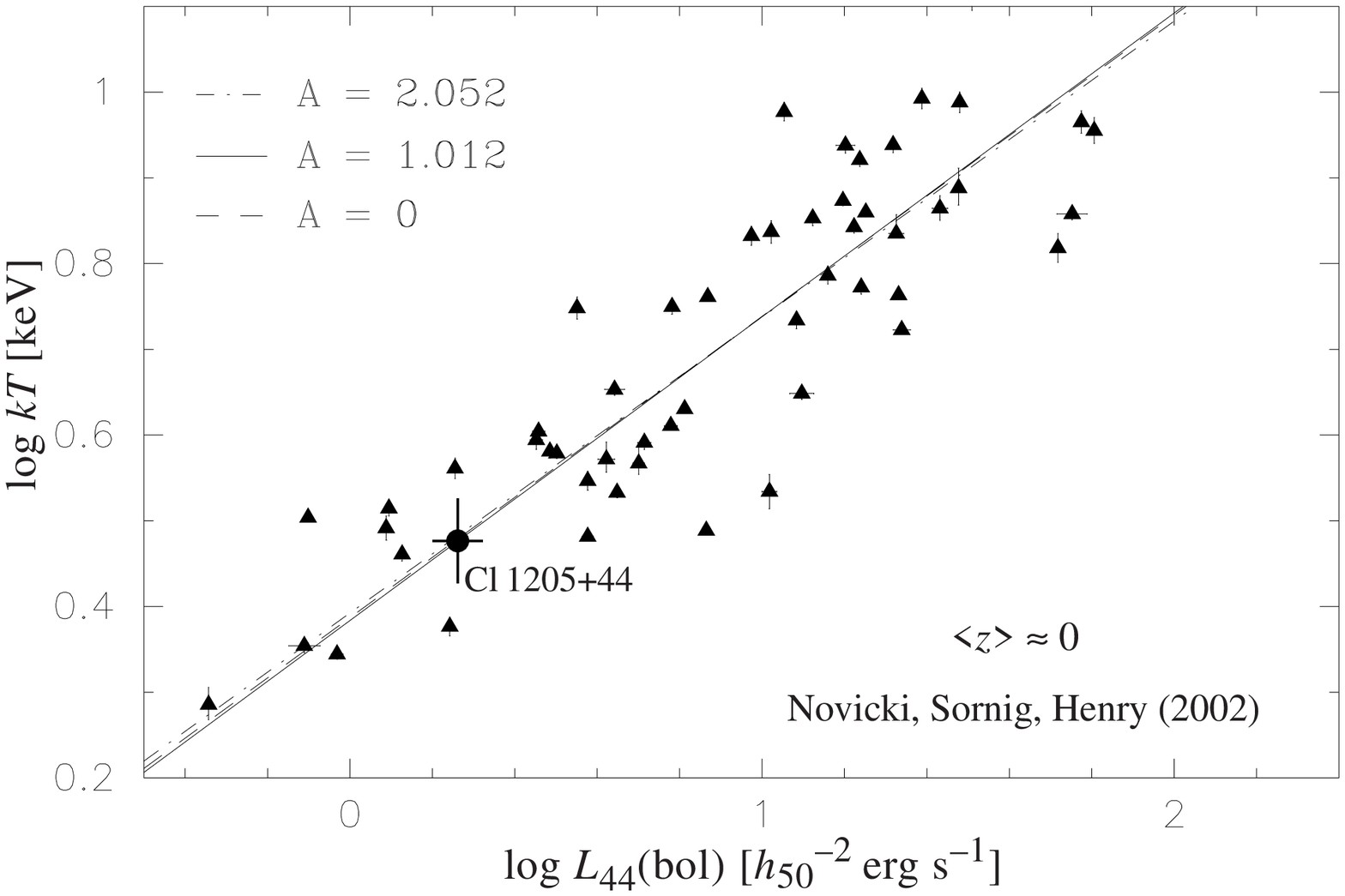,width=4.4in}}
\parbox{4.5 in}{\epsfig{figure=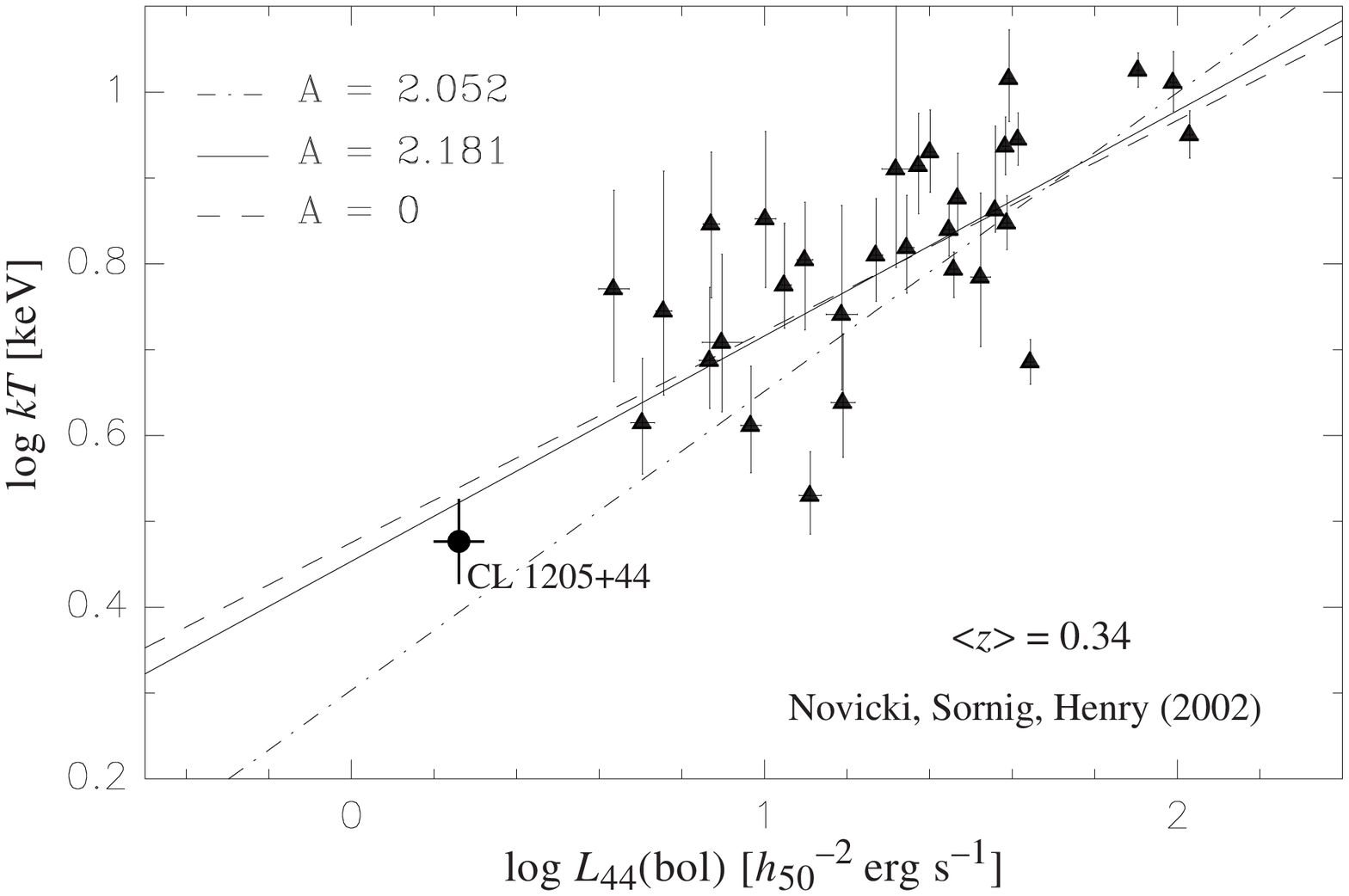,width=4.4in}}
\caption[]{Top: low-redshift data set and solutions ($\Omega_{M}=0.3$, 
$\Omega_{\Lambda} = 0.7$).
${\rm L_X-T_X}$ relation of \cite{Novicki2002} for $z \sim 0$.
Bottom: high-redshift data set and solutions
$z \sim 0.3 $ with  Cl~1205+44 indicated\label{novjones}. 
The dashed lines in both figures correspond
to the no evolution model.}
\end{center}
\end{figure*}

For its measured bolometric luminosity, Cl 1205+44 is hotter than the
best fit to the local $L_{X}$--$T_{X}$ relation based on two fossil
groups (\cite{Jones2003}; see our Figure~\ref{jones23}), but agrees with the
local relation of \cite{Novicki2002}, and the value falls within
$1\sigma$ of the $L_{X}$--$T_{X}$ no-evolution relation derived for
the $\langle z \rangle = 0.34$ sample of \citet{Novicki2002}. For the
purpose of later discussion, we have plotted the values of $kT$ versus
$L_{X}$ for Cl 1205+44 on figures taken from \citeauthor{Jones2003}
and \citeauthor{Novicki2002} in our Figures~\ref{jones23} and
\ref{novjones}.

\subsubsection{X-ray surface brightness fits} 
\begin{figure} \begin{center} 
\parbox{3.25in}{\epsfig{figure=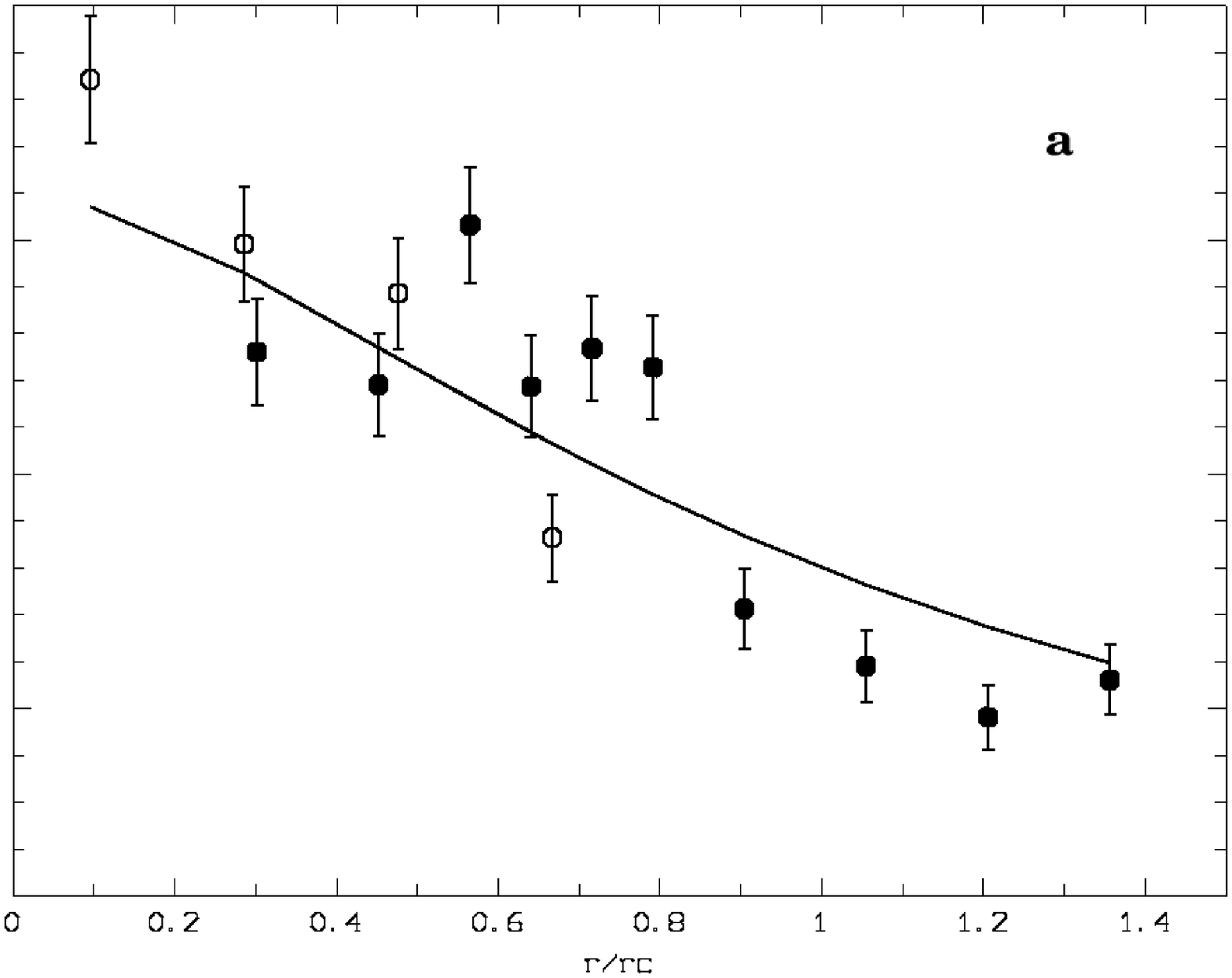,width=3. in,angle=0}} 
\parbox{3.25in}{\epsfig{figure=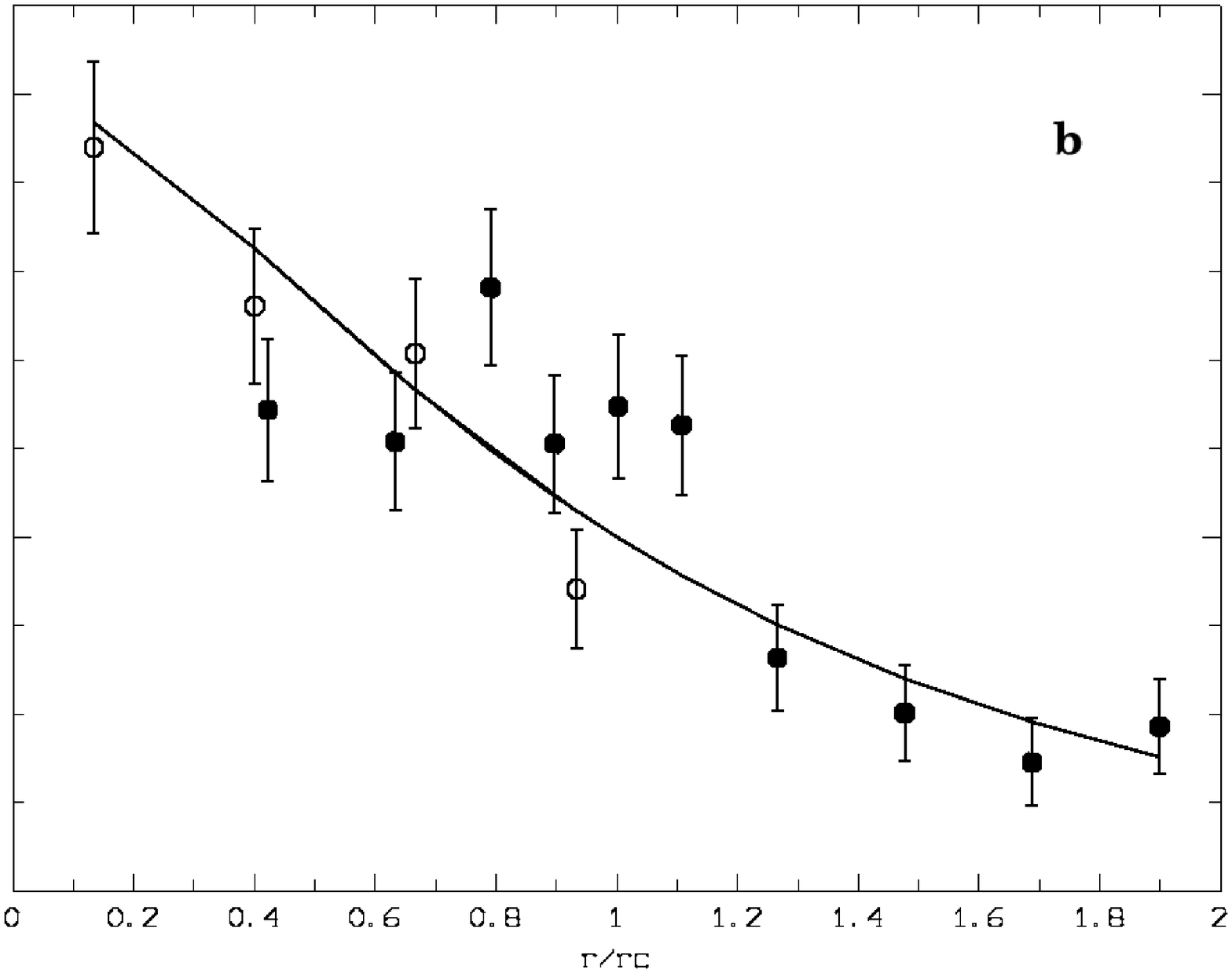,width=3. in,angle=0}}

\caption{\label{profile} Figure 9a: Radial surface brightness profile fit based
on the \xmm\ data (solid) and normalized \cha\ data (open) in
arbitrary units on the Y-axis. The X-axis is in units of core
radius. The best fit (shown a solid curve) and redshift correspond to
$\sim 140$ kpc. Figure 9b: fit for the combined data
set; here the core radius corresponds to $\sim 100 kpc$. See text for details.}
\end{center}
\end{figure}

We carried out a standard $\beta$
surface brightness fit to both the \xmm\ data and the \xmm\ plus \cha\
data combined. We used the 0.5--8 \keV\ bands in both cases. Then with
$I_x(b) \propto [1+[b/r_c]^2]^{-3\beta + 1/2}$
\cite[e.g.][]{Sarazin1986}, where I$_b$ is the surface brightness as
function of projected radius, b and r$_c$ is the core radius. For the
more robust \xmm\ data alone case, we found, using a maximum
likelihood method: $\beta = 0.45 \pm 0.02$, and r$_c = 21\arcsec \pm
3\arcsec$, $1\sigma$ errors.  The fit appears better to the eye (see
Figures \ref{profile} a and b) for the \xmm\ plus \cha\ case, but we
found the results so sensitive to the normalization between the two
data sets and the binning of the data, that we only quote this fit as
a $2\sigma$ lower bound to the core radius = 15\arcsec; the value of
$\beta$ was again 0.45 for this minimum $\chi^2$ fit.

\subsection{Radio Source}  \label{RS}  
\begin{figure}
\begin{center}
\plotone{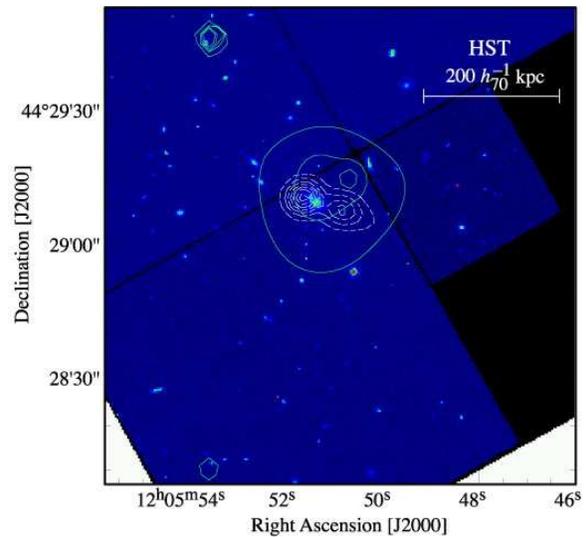}
\caption{\label{firsthst}\cha\ solid contours from the 0.5--8 keV
image (colored green in the electronic version; see also
Figure~\ref{firstcha}) superimposed on the F702W HST image of
Cl~1205+44 along with the dashed contours (colored white in the
electronic version) for the FIRST double-lobed radio source. North is
up, east is to the left.}
\end{center} \end{figure}

\begin{figure} 
\begin{center} 
\plotone{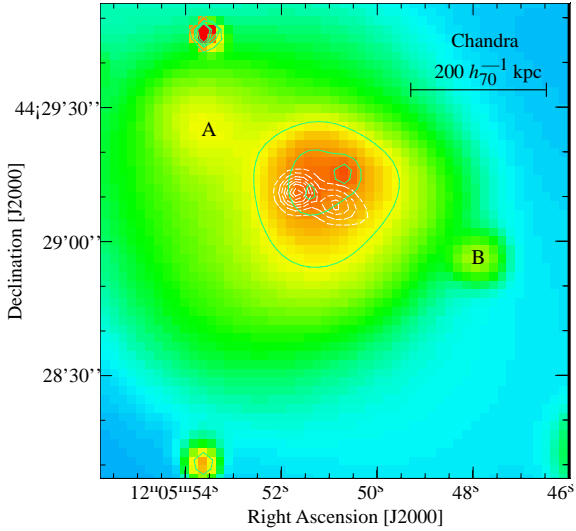}
\caption{\label{firstcha}\cha\ solid contours (green in the electronic
edition) superimposed on a smoothed Chandra 0.5--8 keV image of
Cl~1205+44 along with the dashed contours (white in the electronic
edition) of the FIRST double-lobed radio source. The \cha\ data are
logarithmically spaced and the radio contours linearly spaced. The
\cha\ image (rebinned by a factor 4, i.e., 1 pixel = 2\arcsec) is
adaptively smoothed using the task csmooth from CIAO 3, corrected by
the exposure map. The X-ray contours superimposed on the cluster are
linearly spaced: 4, 8, and $12\sigma$ above the background (that is,
the more or less circular contour is $4\sigma$ above the background,
the two peaks are $12\sigma$ above the background). Then, the contour
levels go geometrically (those over the AGN at the northeast). The
color scale map is, however, logarithmic. The VLA FIRST contours are
linearly spaced, beginning at 0.0035 Jy/beam in steps of 0.0035
Jy/beam.  North is up, east is to the left. The point sources at the
top and bottom of the image to the east of center are line-of-sight
objects that are probably AGNs not associated with Cl~1205+44.  See
the electronic version of the Journal for a color version of this
figure where the \cha\ contours are in green and the radio in white.}
\end{center}
\end{figure}

The central peak within the second
highest contour level in Figure \ref{hstchandra} is located on the D
galaxy.  The D galaxy is also a double-lobed FIRST \citep{Becker1998}
radio source F1205+44 (see Figure~\ref{firsthst}) of total flux =
$56.4\pm 1.7$ mJy at 20 cm \citep{Condon2002}.  These facts are
relevant to the model proposed by \cite{West1994}. He proposed an
anisotropic merger model for the origin of the formation of D
galaxies, and the model predicts that the D-galaxy will be associated
with a powerful radio source, which is the case here.  Furthermore,the
model predicts that the radio lobes will be aligned with the major
axis of the X-ray emission in the cluster. Therefore, we attempted to
determine the major axis of the X-ray emission, which is rather ill
defined.  In order to make a determination of the major axis, we
heavily smoothed\footnote{Note that the heavy smoothing causes the
center of the second highest contour level in the X-ray emission to be
different from that in Figure 3, but the exact position of the X-ray
peak is not important. For, it is extremely unlikely that the D galaxy
would fall so close to the center of this X-ray emission and {\em not}
be associated with the X-ray emission.  The highest contour level has
two peaks, one of which falls on the D galaxy; see the black and white
version of Figure \ref{firsthst} or the color image of Figure
\ref{firstcha}.} the data to produce Figure \ref{firstcha} (see also
Figure \ref{firsthst}). Then, the major axis of the X-ray emission of
Cl 1205+44 can be defined by the line joining points ``A'' and ``B''
in Figure~\ref{firstcha}.  In this case the radio lobe axis is aligned
within 10 degrees of the X-ray axis, which is also consistent with the
model of \citeauthor{West1994}.  However, the optical axis of the D
galaxy is offset by about 30 degrees with respect to the radio lobe
axis and the West model also predicts alignment with the galaxy
distribution. {\em If} there {\em is} a galaxy distribution that
defines a direction, it is the almost north-south line of galaxies
running from galaxies \#7 to \#8 in Figure \ref{melmarked}.  Overall,
then, the data are not consistent with the model of
\citeauthor{West1994}.

\section{Discussion} 
\label{Discussion}

\subsection{Nature of the System}
\label{NOS}

The most compelling reason for calling Cl 1205+44 a fossil group is
that the value of m$_{12}$ within one half the projected virial
radius\footnote{For consistency with \citeauthor{Jones2003} $r_{vir}$
= 70\arcsec\, we used their cosmology, H$_0$ = 50, q$_0$= 0.5; then
the scale is 7.57 kpc/\arcsec, and from their formula for the virial
radius, reproduced in the introduction, r$_{virial}$/2 = 0.53
h$_{50}^{-1}$ Mpc, which corresponds to 70\arcsec.} is very close to
that of the \cite{Jones2003} criterion of being $ > 2$ for {\em z}
$\sim 0$ (for R- or V-band rest frame), i.e. m$_{12}$ ($i'$) =
1.93. In addition, that the value of m$_{12}$ is slightly smaller can be
attributed to Cl 1205+44 being {\em younger} (by up to $\sim 4$ Gyrs)
than $z = 0$ fossil groups; in the \citeauthor{Jones2003}
scenario, the central galaxy grows in brightness with time as it
accretes more galaxies. In older systems the central galaxy will have
had more time to accrete galaxies, and hence be brighter compared to
the remaining ones.

The X-ray emitting AGN we have listed in Table \ref{galaxies} is
brighter than the D galaxy in the $i'$-band, so if it were a cluster
member then Cl 1205+44 would certainly not be a fossil group. However,
we have (without redshifts) two arguments against this AGN being a
group member: (1) its color is significantly bluer than the other
(probable) cluster members in Table \ref{galaxies}; (2) the AGN
becomes the dominant galaxy and then is $\simgreat 2$ (X-ray) core
radii away from the centroid of the extended X-ray emission.

To be a fossil group, the lower limit \cite{Jones2003} place to
L$_{Xbol}$ is $1 \times 10^{42}$ h$_{50}^{-2}$ ergs s$^{-1}$, which
Cl~1205+44 easily meets. Although \citeauthor{Jones2003} do not set an
upper limit to the X-ray temperature or luminosity, the L$_{Xbol}$
(corrected to H$_0$ = 75 for their Figure 3, and our
Figure~\ref{jones23}) point falls in the region where the
\citeauthor{Jones2003} fossil group sample lies. Cl~1205+44 is also
distinguished from ``normal clusters'' in that the L$_{Xbol}$ value is
about a factor of two lower at kT $\sim 3$ keV than the cluster sample
compiled by \cite{Lumb2004} for clusters at {\em z} $\sim 0.4-0.6$. In
contrast, the L$_{Xbol}$-- L$_R$ point lies approximately half way
between the \citeauthor{Jones2003} lines for ``normal'' X-ray bright
groups and fossil groups. On the other hand, the X-ray luminosity and
kT values for Cl~1205+44 are similar to the values \cite{Holden2001}
assigned to an extended X-ray emitting region they simply classified
as a ``cluster.''  In the end, however, the m$_{12}$ value of nearly 2
(within one half the projected r$_{vir}$) and L$_{Xbol} > 10^{42}$
ergs s$^{-1}$ meet the primary Jones et al. criteria, which leads us to
conclude Cl 1205+44 is fossil group, and we will assume it is a fossil
group in what follows.

\subsection{Fossil Group Members}

\begin{figure}
\begin{center}
\includegraphics[width=8cm]{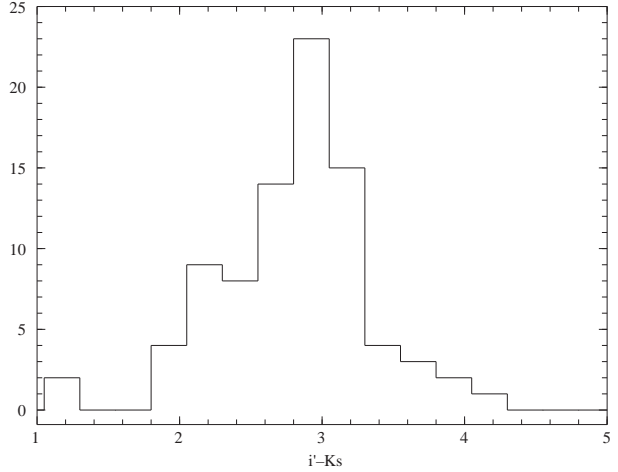} 
\caption{\label{histoik}Histogram of the $i'$-$K_{s}$ colors. Luminous galaxies at
{\em z} $\sim 0.59$ have similar colors
according to a correlation between SDSS and 2MASS photometry.}
\end{center}
\end{figure}

We must make an assumption about galaxy membership, since redshifts
for the specific galaxies are not available to us, and except for the
D galaxy none of those marked in Figure~\ref{melmarked} was measured
by J.~S. Mulchaey (2003, private communication).
The conclusion that these galaxies are
all group members is based on the following facts: their $i'$
magnitudes are all similar except for the D galaxy; their colors are
all similar except for galaxy \#6; their apparent sizes are all
similar, and their average color (see also \S~\ref{GC}) corresponds to
the peak in the histogram of the $i'$--$K_{s}$ distribution shown in
Figure~\ref{histoik}. We will assume therefore, that except for \#6,
all these galaxies are cluster members. Based on their disk-like
morphology, we classified 3 out of the 6 group members as late type
(spiral) galaxies and the other three (including the D galaxy) as
early type (elliptical) galaxies.  We will refer to the galaxy
population and colors in our discussion of the scenarios of this
fossil group formation in \S \ref{GC} and \S \ref{Scenarios} below.

\subsection{Possible Cooling Evolution}

Perhaps the fact that Cl~1205+44 is hotter than the two {\em z} = 0
fossil groups with measured temperatures discussed by \cite{Jones2003}
is due to cooling between {\em z} $\sim 0.6$ and {\em z} $\sim 0$. We
now consider this possibility.  We used the cooling time equation from
\cite{Sarazin1986} for our calculations (see \S \ref{GM} for details).
To derive a lower bound, we used our $\beta$ =0.45, r$_{c}$
=15\arcsec\ model fit and we derived an average cooling time of $6.5$
Gyrs within 1 core radius and $\sim 11$ Gyrs within 2r$_c$.  The time
between {\em z} = 0.59 and {\em z} = 0.2 (the highest redshift in the
sample on which \citeauthor{Jones2003} base their discussion of fossil
group formation and evolution) is only 3 Gyrs. Even the longer time of
4 Gys (the average age of fossil groups in the \citeauthor{Jones2003}
scenario) is less than the 6.5 Gyrs we derived for the average cooling
time within the core. We derive a $2\sigma$ lower bound of $\sim 4$
Gyrs at the very core which could lead to cooling in the center of the
cluster.  It is unlikely, therefore, that the ICM of high redshift
fossil groups is hotter than that of low redshift ones due to cooling
between {\em z} $\sim 0.6$ and {\em z} $\sim 0.0$. We defer a
discussion of the energy input to the ICM until \S\ref{consequences}.

The issue of cooling, gas infall (or the suppression of gas infall), resulting
heating due to gas infall, etc. is a complicated one and has been discussed
extensively in the literature, see for example, some recent works \cite[][and
references therein] {Clarke2004,Kaastra2004,Peterson2004}. The purpose of the
above discussion was not to argue for or against cooling per se, but rather to
determine if it is possible for the gas to have cooled enough between {\em z}
$\sim$ 0.6 and 0.1 to explain the temperature difference between Cl~1205+44
and nearby fossil groups. We have shown that if we assume the simplest
circumstances, i.e. a collisionless gas without a tangled magnetic field, the
gas could just barely cool over this {\em z} $\sim$ 0.6-0.1 time interval.
Then, since heat input is likely in any event, we conclude that the nearby
fossil groups {\em do not} have lower temperatures than Cl~1205+44 because
of cooling. It may just be that hot fossil groups such as Cl~1205+44 are rare
per unit volume but are the ones that are easiest to detect at high {\em z}.

\subsection{The Radio and X-ray Source Relation}\label{TRadio} 

The issue is
whether or not there is evidence for the radio source interacting with the ICM
and producing X-rays via the inverse Compton effect.  The radio  lobes, as we
discussed in \S\ref{RS}, seem to be aligned with the emission that defines the
major axis in the X-ray emission, but there are no radio lobes on either
the FIRST \citep{Becker1998} image or the  NVSS \citep{Condon2002} image that
coincide with the X-ray emission features called ``A'' and ``B'' in \S \ref{RS}. Therefore, this alignment is probably accidental.
Regardless of  whether we assume the alignment is accidental or not, the
contribution of inverse Compton flux from the radio lobes to the 1-10 \keV\
X-ray emission appears negligible as there is no (detectable) X-ray enhancement
associated with the radio lobes.

Based on the ratio of the D galaxy to NGC 1550 (at the center of an
X-ray bright group; \cite{Sun2003} 20 cm fluxes ($56.4\pm 1.7$
mJy/$16.6\pm1.6$ mJy, from the VLA NVSS; \citeauthor{Condon2002},
\citeyear{Condon2002}) and on the measured redshifts, we find that the
radio luminosity of the D galaxy in Cl~1205+44 is about 1.4 $\times
10^4$ times that of NGC 1550. The total radio flux ($\sim 2.5 \times
10^{-14}$ ergs cm$^{-2}$ s $^{-1}$ assuming a flux spectral index of
$-0.7$ and that the flux extends from 0.1 GHz to 10 GHz) is comparable
to the total X-ray flux (and luminosity) of the gas. But if we assume
the radio source is only ``radio active'' \cite[e.g.][and references
therein]{lara2004} for 10$^8$ years, plus that it probably can inject
no more than 10\% of its maximum amount of energy output into the ICM,
this reduces the total energy input to 1\% of the total energy output
of the ICM over 1 Gyr.

Furthermore, there is the lack of a correlation between the radio and the X-ray
emission. Therefore we have no evidence that the radio source is responsible
for extra energy input to the intra cluster medium (ICM) of Cl~1205+44.
Regardless, \cite{Sun2003} suggest that cD galaxies can provide heating to the
ICM via galactic winds over 10 Gyr.

\subsection{Gas Mass and Total Mass}
\label{GM}

In order to compare with previous work, it is interesting to calculate
the gas mass in various ways. Below we give the values based on the
assumption of a core radius of 21\arcsec. The values are approximately
40\% lower if the other best lower limit of 15\arcsec\ is
used. Therefore these results are simply for qualitative comparison
with previous work.

We used the relationships between X-ray surface brightness and mass as
described in \cite{Sarazin1986}.  We assumed the electron density is
1.1 times the proton density and a mean molecular weight of 0.6 for
the gas. Within 100 kpc for the XMM model, we find a gas mass $\sim 8
\times 10^{11}$ M$_{\bigodot}$. Within one core radius the gas mass is
$1.9 \times 10^{12}$ M$_{\bigodot}$. If we assume no temperature
gradient, we derive a dynamical mass (e.g., Sarazin, 1986) of $1
\times 10^{13}$ M$_{\bigodot}$. If we use the \cite{Sun2003} value of
M/L of 5 and our value of L$_R$, we find the total galaxy mass of
Cl~1205+44 is $7.5 \times 10^{12}$ M$_{\bigodot}$ which is over 10
times higher than their value of about $5 \times
10^{11}$ M$_{\bigodot}$ for the group ({\em not} classified as a fossil
group, however) surrounding NGC 1550.  Within 100 kpc, their value of the
gas mass (M$_ {gas}$) of about $3 \times 10^{11}$ M$_{\bigodot}$ is
about 2 times lower than our value for Cl~1205+44. We also calculated
M$_{2500}$ from \cite{Allen2001} \cite[see also][]{Sun2003} who
derived a formula for M$_{2500}$ using a set of X-ray luminous relaxed
clusters. Then, M$_{2500} = 2.7 \times 10^{13}$ M$_{\bigodot} \times
(T/1.37)^{1.5}/E(z) = 3 \times 10^{13}$ M$_{\bigodot}$.  $E(z) = H/H_0
= (\Omega_{r,0}(1+z)^4 +
\Omega_{m,0}(1+z)^3+\Omega_{\Lambda,0}+(1-\Omega_0)(1+z)^2)^{1/2}$;
for $\Omega_{r,0} = 0$ as assumed here, this can be simplified to
$(1+z)(1+z\Omega_{m,0}
+\Omega_{\Lambda,0}(1+z)^2-\Omega_{\Lambda,0})^{(1/2)}.$

\subsection{Galaxy Colors}
\label{GC} 

To compare with other galaxies at this redshift we have used the Sloan
Digital Sky Survey (SDSS)\footnote{http://www.sdss.org/dr2/} and the
2MASS catalog.  For galaxies bright enough to have spectra measured
with the SDSS, we find that the colors of those galaxies at {\em z}
between 0.5 and 0.6 are typical of those in Table~\ref{galaxies}
assumed to be members of CL 1205+44.

The six galaxies for which we have colors that we associate with
Cl~1205+44 are all relatively red, even the late-type galaxies. In
contrast, \cite{BO1984} found a high fraction of {\em blue} galaxies
at this redshift compared to low redshifts. The Butcher-Oemler effect
can be explained by assuming that at 0.6 compared to 0, there is a
higher fraction of galaxies that have just fallen into the cluster and
have not had their gas stripped yet
\cite[e.g.][]{Kauffmann1995}. Hence, the newer (at least spiral)
cluster members tend to be bluer, the higher the redshift.  For
Cl~1205+44, the fossil group could have formed at {\em z} of about 2,
could have aged about 4 Gyrs, and could have had no galaxy infall
since birth. The only galaxy population evolution that has taken place
has been the merging of galaxies into the central D galaxy plus ram
pressure stripping of the gas from the galaxies.  Then, there are no
recent infall spirals and the spirals have the same colors as the
elliptical (under the assumption that all but galaxy \#6 and the AGN
in Table~\ref{galaxies} are cluster members; for, as noted in \S
\ref{NOS} it would be peculiar to have the brightest cluster galaxy so
far removed from the cluster center).  In this hypothesis, the spirals
have had their gas removed via ram pressure stripping, and they have
been cluster members since its formation approximately 4 Gyrs ago.
The key issue is what suppresses galaxy infall for fossil groups
compared to typical groups and clusters.

Fossil groups have probably formed in initially above average over dense
regions (so that the group collapsed early), which, however,  were not
sufficiently rich in matter and galaxies to sustain growth of the group beyond
some point in time. The property of having relatively a low total matter value 
and a negligible blue galaxy fraction is directly related to arriving at an
over density sufficient for collapse earlier than more massive systems. For
example, simulations by \cite{Gao2004} have shown that the galaxy infall rate
and number of galaxies in a cluster is related to the age of the universe when
the cluster formed; clusters that formed earlier in these simulations tend to
have fewer galaxies and less continuous infall (to produce the BO effect) than
the systems that formed later. These simulations predict, then, that
the galaxy to total cluster mass ratio should be lower for the earliest formed
clusters compared to later ones. With a sample of one  fossil group with red
galaxies, however, it is premature to make a comparison  between the data and
the simulations at this level.

\subsection{The D Galaxy}\label{TheD} 

The D galaxy of our system is similar
to the {\em z} = 0.25--0.5 simulated galaxies in \cite{West1994}. Our
value of the $i'$ magnitude of 19.27 converts into a rest frame
absolute R$_J$ magnitude of $-24.1$ (luminosity distance $ = 3.4
\times 10^3$ Mpc, where we used 0.7 for the K correction for an
elliptical at {\em z} = 0.59). This value of $-24.1$ is about 0.5
magnitudes brighter than the brightest cluster galaxy magnitudes
compiled by \cite{Collins2003}. Recent galaxy mergers may be
responsible for this relatively high brightness.

\begin{figure} \plotone{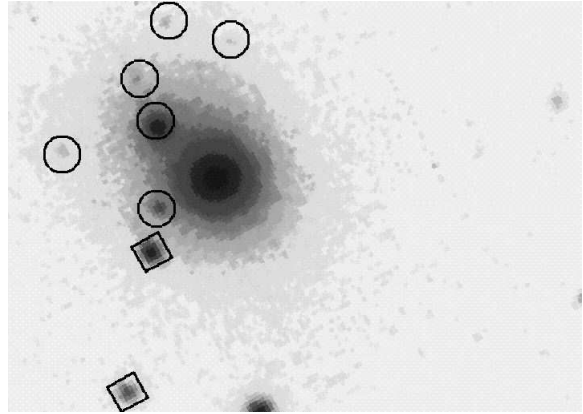} \caption{\label{cD}HST image of the D galaxy
area.  The objects  marked by a square are the galaxies detected in the
$i$'-band data (besides the cD) and the circled objects are resolved in this
HST image but not in the $i'$ band image (not shown). The orientation of the
image is North up and East to the left.} \end{figure}

Note that ongoing minor mergers still appear to be visible on the HST
image (see Figure ~\ref{cD}). These ongoing mergers at a relatively
early stage of this fossil group history do not favor the fossil group
formation scenario of \cite{MZ1999}, who proposed a formation of these
systems with an unusual initial luminosity function.  An alternative
proposal is the evolutionary scenario of fossil groups proposed by
\cite{Borne2000} and \cite{Jones2003}.  This links compact groups of
galaxies and giant elliptical galaxies via an ultra-luminous infrared
galaxy (ULIRG) phase. The D galaxy of our system (which is also a
radio source) has an $i'$-$K_{s}$ color very similar to that of ULIRGs
(from NED\footnote{http://nedwww.ipac.caltech.edu/}) in the 2MASS
survey. This D galaxy could be an ULIRG that is just turning on,
except that the $K_{s}$-band flux falls well below other ULIRGs
\cite[e.g.][]{Yun2004}. Therefore if the D galaxy were formed in the
process suggested by \cite{West1994}, then the ULIRG phase has
probably already passed.

Continuing with the idea of D galaxy evolution in galaxy groups, we refer to
\cite{Collins2003}.  
They show that in poor groups the variance between central
galaxy properties from group to group is quite large (factors of 10 or more) in
terms of total optical output.  Thus, we expect there would be a variation in
the optical properties of the D galaxies in fossil groups such as
Cl~1205+44, {\em unless} there is a direct correlation between the kT
value and the D galaxy stellar content. Only a high redshift ($\simgreat
0.6$) survey of fossil groups will be able to shed light on the variance of
initial conditions and a survey in general is needed to shed light on the
relationship between kT and the D galaxy.

\subsection{Scenarios} \label{Scenarios}

To summarize a formation scenario for Cl 1205+44: There was an initial
potential well of dark matter. Energy injection occurred via some
process or processes such as perhaps supernovae, radio galaxies, or
ULIRGs.  This process heated the gas as galaxies fell into the
potential well at a redshift of about 2.  The galaxies passed through
the ICM several times and the gas and dust were swept out of the
spirals. This ram pressure stripping suppressed continuous star
formation and left the spirals to become as red as the ellipticals.
The system formed in a relative void of galaxies so that there was no
continuous infall of galaxies to keep the system fed with blue spiral
field galaxies.  The central galaxy grew to become a relatively large
D galaxy by merging with other galaxies to form a m$_{12}$ value
$\simgreat 2$. The entire evolutionary sequence took about 4 Gyrs.

The above fits with the \cite{Jones2000,Jones2003} scenario, which
assumes that the time it takes for the merging of galaxies to form a
bright central galaxy is approximately 4 Gyrs and no continuous field
galaxy infall has taken place.  The fact that the D galaxy in
Cl~1205+44 appears to have nearly completed the merging process and is
approximately 4 Gyrs younger than {\em z} $\simeq 0$ fossil groups,
implies that (at least some) fossil groups at {\em z = 0} are much
older than 4 Gyrs.

\subsubsection{Consequences} 
\label{consequences} 

Energy to heat the ICM above and beyond gravitational infall results in an
entropy (defined here as the pseudo entropy kTn$_e^{-2/3}$; where n$_e$ is the
number of electrons per cm$^{-3}$ and kT is in \keV) floor. The initial energy
injection and its consequences were discussed in detail by \cite{Babul2002}.
\citeauthor{Babul2002} made predictions of initial energy input which they
compared with the data. The two fossil groups found by \cite{Jones2003} with
strong enough signal to measure a temperature were consistent with an entropy
floor of 100 \keV\ cm$^2$. In contrast, Cl~1205+44 lies a factor of two or
more above the 100 \keV\ cm$^{2}$ line and instead is nearly on the line
occupied by normal rich clusters and normal groups (initial energy input 427
\keV\ cm$^{2}$; see Figure \ref{jones23}). This implies that either Cl~1205+44
is really not a fossil group, contrary to our classification, or that it is
unlike the nearby fossil groups, and that we were able to detect Cl~1205+44
precisely because it is much more luminous than low {\em z} fossil groups.
This could be due to the result of a relatively high (compared to the average
fossil group) entropy floor input in the ICM, or indicate that Cl~1205+44
simply is more massive that other fossil groups.

Opposed to the model of \cite{Babul2002}, we find a much lower value
for the pseudo entropy = Tn$_e^{-2/3}$ \keV\ cm$^{2}$ of 85 \keV\
cm$^{2}$. We derived this value at r = 0.1r$_{200}$ = 0.1r$_{virial}$
from our $\beta$ fit (r$_c$ = 21\arcsec), where we estimate an average
kT = 3 \keV and an electron density at the core to be $\sim 8.1 \times
10^{-3}$ cm$^{-3}$, i.e. $6.5 \times 10^{-3}$ cm$^{-3}$ at
0.1r$_{virial}$. This implies some refinement to the
\citeauthor{Babul2002} model such as continuous rather than impulsive
heating (pre-heating) at formation. Note that our entropy for a 3 \keV\ (the
temperature of the ICM of Cl~1205+44) temperature is lower than was
found for 3 \keV\ virialized objects (including clusters and groups)
at {\em z} $\simless 0.2$, by \cite{Ponman2003}.

The fact that the entropy we find is lower than that found for more
nearby clusters and groups at the same temperature is consistent with
the ``negative'' entropy evolution suggested by
\cite{Maughan2004}. \citeauthor{Maughan2004} proposed a form
E$(z)^{-4/3}$ based on self-similar scaling and on the evolution of
the critical density of the Universe with {\em z}.  The {\em
scaled-by-temperature} (i.e. divided by kT) entropies for similar
temperature objects found by \cite{Ponman2003} at the same radius
(0.1r$_{vir}$) have values of about 75 cm$^2$, which if scaled from
{\em z} = 0 to z = 0.5915 by E$(z)^{-4/3}$ correspond to 49 cm$^2$,
which is within a factor of two of the value for Cl~1205+44 = 28
cm$^2$ at 0.1r$_{vir}$.  These results then appear to be more
consistent with continuous rather than impulsive initial heating to
the ICM.

If there were some initial energy injection at {\em z} = 2 
(whatever its cause),  then this injection could affect the high order portion
of the CMB via the SZ effect from clusters and groups of galaxies. The
possibility of this energy input and its effect on SZ measurements has been
discussed by \cite{Lin2004} in the context of deriving cosmological parameters
based on the power spectrum of the CMB.  Their impulsive heating model requires
energy injection at {\em z} = 2. They fit their models to the L$_X$-T$_X$
normal cluster and group data that are shown here in Figure ~\ref{jones23}, and
from this we see that Cl~1205+44 is consistent with the
\citeauthor{Lin2004} model \citep[see also][]{Borgani2004}. The fact that
preheating occurred in the \citeauthor{Lin2004} model near {\em z} $\sim 2$ is
consistent with our suggestion that at least some fossil groups, such as Cl
1205+44, formed at {\em z} $\sim 2$. However, if continuous heating models
can also be shown to fit the data, then the amount of initial impulsive heating
suggested by \citeauthor{Lin2004} will not be as large as they suggested, and
the impact on the high order CMB observations will be less.

\section{Summary and Conclusions}

We have found, via a ROSAT survey of archival data, a group of
galaxies with a relatively red population. HYPERZ calculations based
on the $i'$- and $K_{s}$-band data suggested the system might have {\em z}
$\simgreat 1$.  Subsequent redshift work found that the redshift is =
0.5915, and \cha\ plus \xmm\ observations confirmed the existence of
X-ray emitting gas in the group.  We classify this group as a fossil
group based on its m$_{12}$ in $i'$ and on the fact that its X-ray
luminosity exceeds $ 1 \times 10^{42} h_{50}^{-2}$ ergs s$^{-1}$. This
makes it by far the highest redshift fossil group yet reported.  The
temperature for the ICM is abnormally high compared to two $z \sim 0$
fossil groups with a similar L$_{Xbol}$, and the L$_{Xbol}$--L$_R$
point lies approximately half way between the best fit lines for
normal and fossil groups \cite{Jones2003}; Cl~1205+44 has a
kT--L$_{Xbol}$ value comparable to another {\em z} $\sim 0.6$ system
that was classified as a cluster (RX J0848+4456;
\citeauthor{Holden2001}, 2001).  But, although there are some
similarities with normal groups and clusters, we have classified it as
a fossil group based on the primary criteria of m$_{12}$ $\simgreat 2$
in R- or V-band rest frame and an X-ray luminosity $ \geq 1 \times
10^{42} h_{50}^{-2}$

Our main conclusion is that Cl~1205+44 is a fossil group. A formation scenario
is that the group formed in a peak density region of relatively low total mass
at {\em z} $\sim 2$. The formation regions did not have enough matter to
sustain continuous infall of galaxies and gas, and such a scenario is
consistent with the simulation carried out by \cite{Gao2004}. Beyond the
formation of a central dominant galaxy via accretion, fossil groups appear
old, i.e., fossil-like precisely because they have had little or no evolution
in terms of galaxy infall since their inception. And, some fossil groups could
be considerably older than 4 Gyrs. No evolution of the galaxy population (at
least in the case of Cl 1205+44) is implied because of the dearth of blue
spirals, the kT-L$_{Xbol}$ value point on the no-evolution tracks for rich
clusters of galaxies \citep{Novicki2002}, and our cooling calculations of the
IGM are consistent with (but cannot exclude) no significant cooling of CL
1205+44 in 3 Gyrs.

Some preheating of the gas in groups and clusters via supernovae or an
active galaxy phase is likely to have occurred even in fossil groups
(or normal groups) as evidenced by the relatively high temperature of
CL 1205+44, but it also appears likely that there is continuous rather
than impulsive heating at group formation, which results in
``negative'' evolution of the pseudo entropy
\cite[e.g.][]{Maughan2004}. However, as the kT versus L$_{Xbol}$ of
Cl~1205+44, fits within $1\sigma$ the no-evolution with {\em z} of
\cite{Novicki2002}, the negative evolution for entropy and the
no-evolution models for kT versus L$_{Xbol}$ (and galaxy population)
may need to be fine tuned to be made consistent with each other.

Also, the amount of preheating has implications for the interpretation
of high order measurements of the CMB. Our data are consistent with
``negative'' evolution of the pseudo entropy which implies, on one
hand, a lesser impact of the SZ effect from high {\em z} clusters than
suggested by \cite{Lin2004}. On the other hand, that the kT versus
L$_{Xbol}$ value for our data and many other clusters is consistent
with no evolution, suggests that perhaps a significant faction of ICM
heating did take place near {\em z =2}, and hence does have a
significant effect on the high order terms of the CMB.

The discovery and X-ray and optical measurements of still more  {\em z} $ \geq
0.6$  clusters and fossil groups are needed to develop enough statics to
address the issues of evolution,
preheating, and the relationship of fossil group (and cluster) formation to
their total mass and galaxy colors.  

\begin{acknowledgements} We thank M. Smith for reducing the $K_{s}$-band
data. We thank J. S. Mulchaey for useful discussions and for
giving us information prior to publication on the optical spectra of galaxies
in Cl~1205+44. We thank A. Kravtsov for useful discussions about simulations of
cluster formation. We thank D. Neumann for useful discussions, and we thank the referee for many useful comments which
greatly improved the paper.  This work was supported in part by NASA grants
G03-4156X (subcontract from Smithsonian Astrophysical Observatory) and
NAG5-13556. Also,  NASA LTSA award NAG5-11634 (KS \& AKR) and NASA XMM award
NAG5-12999 (KS) supported this project. We thank the groups and many people
responsible for the successful launch and operation of \cha\ and \xmm\ and that
helped set up our observations. This research has made use of the NASA/IPAC
Extragalactic Database (NED) which is operated by the Jet Propulsion
Laboratory, California Institute of Technology, under contract with the
National Aeronautics and Space Administration.  The i$'$ and $K_{s}$ data were based
on observations obtained with the Apache Point Observatory 3.5-meter telescope,
which is owned and operated by the Astrophysical Research Consortium.  
\end{acknowledgements}

\bibliography{cluster_ref} 
\bibliographystyle{apj}
\clearpage

\clearpage


\clearpage

\clearpage 


\clearpage  

\clearpage
\begin{deluxetable}{llcccc}
\tablewidth{0pt}
\tablecaption{\label{galaxies} Probable Group Members}
\tablehead{ \colhead{ID(Class)$^a$} & \colhead{Ra } & \colhead{Dec} & \colhead{F702w} &  \colhead{i$'$}  
& \colhead{$K_{s}$} }
\startdata 
1(AGN)$^b$ & 181.474 &  44.4960 & 19.27 & 18.23 & 16.30 \\ 
2(S) & 181.472 &  44.4938 & 21.04 & 20.80 & 17.54 \\ 
3(E) & 181.466 &  44.4923 & 21.38 & 21.19 & 18.05 \\ 
4(S) & 181.470 &  44.4885 & 20.73 & 20.43 & 17.29 \\ 
5(S) & 181.460 &  44.4885 & 21.84 & 20.65 & 17.63 \\ 
6(S)$^c$ & 181.469 &  44.4873 & 21.01 & 20.99 & 18.57 \\ 
7(D) & 181.464 &  44.4860 & 19.27 & 18.50 & 15.44 \\
8(E) & 181.465 &  44.4812 & 21.22 & 20.85 & 17.85 \\
\enddata
                                                                                                                           
\tablenotetext{a}{An identification number (ID), and classification as either active galactic nucleus
(AGN), spiral (S) or elliptical E).}
\tablenotetext{b}{Assumed not to be a group member, but included for comparison
and completeness. This object is a point X-ray source as can be in
Figure~\ref{firstcha}. }
\tablenotetext{c}{Possibly not to  a group member, see text.}

\end{deluxetable}


\begin{deluxetable}{lcccc}
\tablewidth{0pt}
\tablecaption{\label{counts}Photon count per detector}
\tablehead{ \colhead{\ } & \colhead{MOS1} & \colhead{MOS2} &  \colhead{PN}  
& \colhead{ACIS} }
\startdata 
Total$^a$ &   180 &  153 &   458 & 465 \\
Cluster  & 147  & 117   & 381  & 295\\
\enddata
                                                                                                                            
\tablenotetext{a}{Total means cluster+background. These counts were 
used for the fitting of the spectra from the cleaned event
files and within the energy ranges described in \S 2.3.3.}
\end{deluxetable}

\begin{deluxetable}{cccc}
\tablewidth{0pt}
\tablecaption{\label{tbl:mekalfit}Thermal plasma best fit parameters
}

\tablehead{
\colhead{$kT$}   & \colhead{ $Z$}          & \colhead{$N_{\rm H}$} & 
\colhead{$\chi^{2}/$d.o.f.} \\
\colhead{[keV]}  & \colhead{[$Z_{\odot}$]} & \colhead{[$10^{20}$cm$^{-2}$]}
} 
\startdata
$3.0_{-0.4}^{+0.5}$ & $0.30_{-0.24}^{+0.30}$ & $<3.77$     & 48.02/46 \\
$3.0_{-0.3}^{+0.3}$ & $0.30_{-0.23}^{+0.30}$ & $1.27^{a}$  & 48.03/47 \\
$3.0_{-0.4}^{+0.5}$ & $0.30^{a}$             & $<3.53$     & 48.02/47 \\
$3.0_{-0.3}^{+0.3}$ & $0.30^{a}$             & $1.27^{a}$  & 48.03/48 \\

\enddata

\tablenotetext{a}{fixed value.}
\tablenotetext{\ }{Temperature ($kT$), 
metallicity ($Z$), and hydrogen column density ($N_{\rm H}$); d.o.f. are 
the degrees of freedom for a given spectral fit.}
\end{deluxetable}

\begin{deluxetable}{ccccccc}
\tablewidth{0pt}
\tablecaption{\label{tab:sum}Main characteristics of Cl~1205+44}
\tablehead{
\colhead{{\em z}} & \colhead{{\rm Alpha$^a$}} & \colhead{{\rm Delta$^a$}} & 
\colhead{X-ray}  & 
\colhead{{\rm $L_{X}$}} & \colhead{{\rm $T_{X}$}}  & \colhead{{\rm $L_{opt}$} (F702W)}\\  

 & \colhead{deg} & \colhead{deg} & \colhead{diameter} & \colhead{$h_{70}^{-2} 10^{43}$ erg s$^{-1}$} 
 & \colhead{keV} & \colhead{$10^{11}$ (${\rm L_\odot})$} }   
\startdata
0.59  &  181.4641 & 44.4860 & 40\arcsec\  & 9.2$\pm$0.7 & 3.0$\pm$0.3 & 1.5 \\
\enddata
\tablenotetext{a}{J2000}
\end{deluxetable}

\end{document}